\documentclass[9pt,twocolumn,twoside]{opticajnl}
\journal{opticajournal} %

\setboolean{shortarticle}{false}

\usepackage{lineno}

\usepackage{caption}
\usepackage{subcaption}
\usepackage{placeins}
\usepackage{tikz}

\usepackage{color}
\usepackage{nicefrac}
\usepackage{xspace}

\newcommand{\HG}[1]{\ensuremath{\text{HG}_{#1}}}
\newcommand{\LG}[1]{\ensuremath{\text{LG}_{#1}}}
\newcommand{\hgket}[3]{\left|\HG{#1},{#2};{#3}\right\rangle }
\newcommand{\hgstate}[3]{\ensuremath\HG{#1; #2}^{#3}}
\renewcommand{\hgket}{\hgstate}

\newcommand{\Gaussian}{Gaussian\xspace} %
\newcommand{\Figure}{Figure} %
\newcommand{\Equation}{Equation} %

\newcommand{\citeg}[1]{(e.g. \cite{#1})}%
\newcommand{\mRtHz}{\ensuremath{\text{m}/\sqrt{\text{Hz}}}}

\newcommand{\tagrepeat}[1]{\tag{\ref{#1} repeated}}

\usepackage{ulem}

\DeclareMathOperator\erf{erf}
\newcommand{\signal}[1]{\textbf{#1}}

\newcommand{\uwa}{Department of Physics, University of Western Australia, Crawley, 6009, Australia}
\newcommand{\ozgrav}{OzGrav, Australian Research Council Centre of Excellence for Gravitational Wave Discovery, Australia}
\newcommand{\ua}{School of Physical Sciences, University of Adelaide, Adelaide, 5005, Australia}
\newcommand{\hamburg}{Institut f\"ur Quantenphysik und Zentrum f\"ur Optische Quantentechnologien der Universit\"at Hamburg, Luruper Chaussee 149, 22761 Hamburg, Germany}
\newcommand{\louvian}{Centre for Cosmology, Particle Physics and Phenomenology, Universit\'e Catholique de Louvain, Louvain-La-Neuve, B-1348, Belgium}
\newcommand{\rome}{INFN, Sezione di Roma Tor Vergata, I-00133 Roma, Italy}
\newcommand{\nikhef}{Nikhef, Science Park 105, 1098 XG Amsterdam, The Netherlands}
\newcommand{\uniroma}{University of Rome Tor Vergata, I-00133 Roma, Italy}

\title{Transverse Mode Control in Quantum Enhanced Interferometers: A Review and Recommendations for a New Generation}

\author[*,1,2]{Aaron~W.~Goodwin-Jones} %
\author[3]{Ricardo~Cabrita} %
\author[4]{Mikhail~Korobko} %
\author[7]{Martin~van~Beuzekom} %
\author[1,5]{Daniel~D.~Brown} %
\author[8,6]{Viviana~Fafone} %
\author[3]{Joris~van~Heijningen} %
\author[6]{Alessio~Rocchi} %
\author[1,5]{Mitchell~G.~Schiworski} %
\author[7]{Matteo~Tacca} %

\affil[1]{\ozgrav}
\affil[2]{\uwa}
\affil[3]{\louvian}
\affil[4]{\hamburg}
\affil[5]{\ua}
\affil[6]{\rome}
\affil[7]{\nikhef}
\affil[8]{\uniroma}
\affil[*]{aaron.jones@ligo.org, \url{https://aaronwjones.com}} %

\begin{abstract}
Adaptive optics has made significant advancement over the past decade, becoming the essential technology in a wide variety of applications, particularly in the realm of quantum optics. One key area of impact is gravitational-wave detection, where quantum correlations are distributed over kilometer-long distances by beams with hundreds of kilowatts of optical power. Decades of development were required to develop robust and stable techniques to sense mismatches between the Gaussian beams and the resonators, all while maintaining the quantum correlations. Here we summarize the crucial advancements in transverse mode control required for gravitational-wave detection. As we look towards the advanced designs of future detectors, we highlight key challenges and offer recommendations for the design of these instruments. We conclude the review with a discussion of the broader application of adaptive optics in quantum technologies: communication, computation, imaging and sensing.
\end{abstract}

\setboolean{displaycopyright}{false} %

\begin{document}
\maketitle

\section{Introduction}
Adaptive optics (AO) refers to any technology used to dynamically correct optical aberrations. Since its inception in the late 1970s~\cite{Hardy_Active_Optics}, AO has played a pivotal role in various scientific and commercial applications. Possibly, the most famous example of AO is the correction of atmospheric turbulence for astronomical telescopes~\cite{Babcock_1958, Glindemann2000}. The past two decades have seen the rapid development and commercialisation of AO technology. As a result, AO is now also used in a wide variety of fields including Free Space Optical Communications (FSOC)~\cite{Stotts21,Facebook_Air_Ground_Link, Karpathakis23}, precision measurements~\cite{Rocchi11,Brooks16}, microscopy~\cite{Ji2010} and biological imaging~\cite{Pircher17}.

Terrestrial gravitational-wave detectors (GWDs)~\cite{ObservingScenarios} are exquisite optical systems. They use high optical powers and nearly perfect resonators to achieve a quantum-noise limited sensitivity across their observing band. These km-scale interferometers are able to detect differential length changes of about $10^{-20}\,\mRtHz$~\cite{O3_instr_paper}. 

Active optical techniques were considered for GWDs as early as 1984~\cite{anderson84}. In 2003, researchers began attempting to directly translate astronomical AO techniques to \Gaussian beams and GWDs~\cite{Calloni03}. However, deploying active optics in quantum enhanced instruments poses numerous technical challenges. Particular issues include back-scatter, polarisation, vacuum and low frequency stability. In the past 20 years, the GW community along with many other independent researchers have developed bespoke sensing and actuation schemes which integrate adaptive optics into the precision engineered system. The resulting detectors are able to significantly suppress quantum noise at audio frequencies~\cite{geo600_6db, Tse2019, Acernese2019}. This achievement has required decades of development of AO. 

To date, uptake of AO systems has generally been constrained to incoherent sources of light~\citeg{Pircher17,Ji2010,Glindemann2000}. In contrast, GWDs, FSOC and many quantum optics applications generally use coherent light and spherical mirrors. The eigenmodes of these optical systems are \Gaussian modes~\cite{Fox61, KogelnikandLi66, Siegman, Bond2017}. We provide a brief introductory summary of \Gaussian modes in Appendix~\ref{app:formalism}. Throughout this review we generally consider the Hermite-Gaussian (HG) modes due to residual astigmatism in GWDs~\cite{Sorazu13, Bond11}. When used, Laguerre Gaussian (LG) modes are defined with the radial index, $p$, following the azimuthal index, $l$, \LG{pl}. In \Gaussian optics, the transverse properties of a laser beam are described by a complex beam parameter,
\begin{align}
    q(z) = iz_R + (z - z_0) \label{eq:q}
\end{align}
where $z_R$ is a scaling parameter describing how quickly the beam expands and $z_0$ is a positioning parameter describing where the beam radius reaches a minimum. The problem of ensuring that the complex beam parameter of the incoming light matches the complex beam parameter of the resonator is referred to as mode matching. Additionally, the HG modes describe only perfect spherical mirrors. However, any real optical system has surface figure errors and clipping. These effects shift the mode basis away from the analytical solutions given by HG-based model~\cite{Ciobanu2021}. Instead, numerical models such as the Linear Canonical Transform (LCT) or Fast Fourier Transform (FFT) must be used to compute the mode basis~\cite{Ciobanu2021}. We refer to this as mode basis degradation.

Despite significant development in AO, mode mismatch accounts for a substantial fraction of the optical loss budget in today's GWDs~\cite{Tse2019,Acernese2019}, thus limiting the reduction in quantum noise. Future detectors, envisioned to start operation in $\sim2035$~\cite{design_study_update_et,ce_horizon_study,nemo}, will require higher levels of quantum-noise suppression, which will require a reduction in the static optical mismatch. Furthermore, they plan to operate with significantly increased optical powers. These optical powers introduce thermal transients, thus requiring better control of the dynamic mode mismatch. Therefore, it is required that wavefront control improves substantially over the next 10 years.

This paper is a historical review of the key developments in the field. In the interests of brevity, we assume some familiarity with GWDs, a complete introduction to the field, including all field-specific nomenclature, can be found in Appendix~\ref{sec:quantum_enhanced_ifo}. Furthermore, a more detailed discussion of quantum noise and how it links to mode matching can be found in Appendix~\ref{sec:quantum_enhanced}. Mismatch sensing schemes for \Gaussian beams are summarised in section \ref{sec:sensors}. Then, we summarise in section \ref{sec:actuators} the custom actuators designed to meet the demanding noise tolerances in opto-mechanics. In section \ref{sec:sites}, we summarise the installed AO at the sites.

As quantum techniques begin to be used more generally, it will be critical to match both free space and fibre optical modes between resonators with minimal loss. In section \ref{sec:the_future}, we make recommendations on the use of the developed technology within GWD and speculate more broadly on applications within quantum information science. 
\subsection{70 Years of adaptive optics}
\label{sec:years_of_adaptive_optics}
AO is the process by which optical aberrations are corrected by active elements. The most simple example is the use of a fast steering mirror (FSM) to maintain the alignment of a laser beam to a quadrant photodiode (QPD) in the presence of seismic or atmospheric turbulence. The technique is well developed with applications in astronomy, vision science, microscopy and FSOC. 

The first proposal for AO was from Babcock in 1953~\cite{Babcock_1958}. Babcock suggested using a knife edge (see appendix \ref{sec:beam_profiling}) to interrogate the beam and feed it back to an Eidophor (a precursor to modern spatial light modulators) which would spatially modulate the phase of the beam. Initially, the technique was focused on astronomical and defense applications to correct for atmospheric turbulence. In subsequent years several review articles were published~\cite{Hardy_Active_Optics,Glindemann2000}. One key development in AO is to use the Zernike polynomials~\cite{F.Zernike1934} to describe the wavefront deformations. These wavefront deformations may be determined using a Hartmann sensor~\citeg{Brooks07} with incoherent light from a distant stellar source. This can be corrected using a deformable mirror~\cite{RoddierBook}. In astronomy, laser guide stars~\cite{Dorgeville16} are used to provide point spread functions to the telescope. For a thorough treatment, please see~\cite{RoddierBook}. 

Since the development of AO for astronomy, several other fields have made use of the technology. For example, in vision science adaptive optics can be used to mitigate time-dependant imperfections in the lens of the eye and allow high resolution photographs of the retina~\cite{Pircher17}. In astronomy a distinction is made between \textit{active optics} and \textit{adaptive optics}. Active optics refers to translational and rotational control of mirrors, whereas adaptive optics refers to higher-order effects. In microscopy, a similar approach can be used to overcome lensing by time-varying flows in the imaged medium~\cite{Ji2010}. A comparison and review of the work in the first three fields is presented in~\cite{Hampson21}. Recent work developing free space optical communication links has to overcome atmospheric turbulence~\cite{Stotts21}.

\section{Proposed mismatch sensing schemes}
\label{sec:sensors}

Several optical mode sensing schemes have been proposed that meet the requirements of GWDs. We group these into indirect methods---which image the effects of thermal transients, direct methods---which directly measure the mode matching between resonators, and mode decomposition---which attempts to decompose the beam and identifying the mode structure and mode basis. However, prior to discussing mode sensing specifically, we will comment on the design considerations which minimise the need for modal actuation. 

\subsection{Design considerations}
\label{sec:design_considerations} 
The Rayleigh range of the eigenmode in the arms of LIGO detector is $\sim400$\,m~\cite{AdvancedLIGO15}. As such, the mode matching between the input optics and arms cannot be corrected by distance optimization alone. Furthermore, given the size of the beams, it is not possible to profile the beam directly. Therefore, the designs of these telescopes~\cite{Arain08, Granata10a, Rowlinson17} are critical to avoid static mismatches in the interferometer. 

The designs of such telescopes typically use off-axis spherical or parabolic mirrors to avoid back-reflections from imperfect anti-reflective coatings. Such telescopes can be designed using spherical mirrors with angles that minimally couple astigmatism~\cite{Hello96}. A thorough tolerance analysis is then undertaken in simulation to minimize the sensitivity to possible polishing errors. This must be done while also ensuring that the cavities are geometrically stable, thus strongly enforcing the modal basis and facilitating alignment control. 

An alternative option, pursued by Advanced Virgo, is to use marginally stable recycling cavities. In this case, the high magnification telescopes are shifted to the input optics~\cite{Buy17}. The design of the input and output optics requires a similar tolerance analysis~\cite{T1300941,T1900649}.

To date, GWDs have been designed to use spherical mirrors, thus enforcing the HG mode basis. However, a series of papers~\cite{DAmbrosio03, Dambrosio04} followed by preprints~\cite{Oshaughnessy04,ambrosioX04} considered the use of non-spherical mirrors to reduce thermal noise. Recently, non-spherical mirrors have been proposed once more~\cite{Richardson22}, to shift the resonance of higher-order modes away from degeneracy with the \HG{00} mode. The recycling cavities can then be designed by constructing an appropriate cost function and using Monte Carlo based simulations~\cite{Richardson22}.

\subsection{Indirect mismatch-sensing techniques}
\label{sec:indirect}

\begin{figure}
    \centering
    \includegraphics[width=0.8\linewidth]{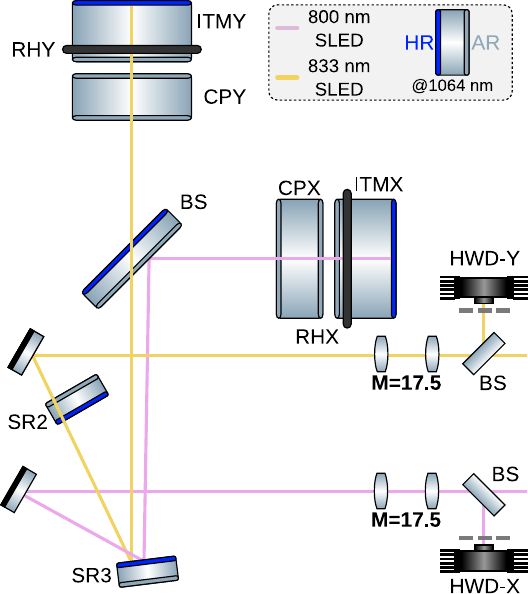}
    \caption{Schematic representation of the Advanced LIGO wavefront sensors. SLEDs inject light at 800\,nm (X-Arm) and 833\,nm (Y-arm). The AR coating on the BS reflects the 800\,nm beam and transmits the 833\,nm beam allowing for independent measurements of X \& Y test masses. Ring Heater X (RHX) and RHY are actuators, see section~\ref{sec:ring_heaters} for details. Adapted from \cite{Brooks16} with permission.}
    \label{fig:Brooks16_Fig6}
\end{figure}

One of the crucial developments is an ultra-low noise Hartmann Wavefront Sensor (HWS).
These devices differ from the Shack-Hartmannn sensors~\citeg{RoddierBook}, common in other AO applications, as the micro-lens array is replaced with a plate of uniformly spaced pinholes, though the operating principle is the same.
Forgoing the lenslet array allows for higher sensitivity as the effect of aberrations from imperfections in the lenses is removed, but also results in greatly reduced light collecting efficiency \cite{Chernyshov:05,Brooks07}.
Both Advanced LIGO and Advanced Virgo use the HWS described in~\cite{Brooks07}.
In the GWD implementation, probe beams of incoherent light are retro-reflected through transmissive optics and onto the HWS, as illustrated in \Figure~\ref{fig:Brooks16_Fig6}.
The measured distortions of these probe beams are, hence, related to distortions in the actual test masses themselves. The HWS can be calibrated to the cold mirror surface maps, which are measured independently. As the mirror thermally deforms, the HWS can then provide information on the current absolute mirror surface to a high precision. 

The Hartmann sensor is an indirect method of sensing mismatch. The result contains information on the wavefront deformations, but contains no information on the overlap between the cavity eigenmode wavefronts and the injected wavefronts. In the case of gravitational wave detectors, the lensing which degrades mode matching is typically caused by temperature gradients due to optical absorption. Each test mass supports a number of vibrational eigenmodes, the spectrum of which depends on temperature, and thus another indirect mode sensing method is to track these eigenmode frequencies. This idea was first reported in~\cite{Blair20}. 

\subsection{Direct mismatch sensing techniques}
\label{sec:direct_mm}
The most common way of inferring the laser beam parameter directly is to profile the beam. Common approaches include: the knife edge method, the chopper wheel, the scanning slit and the camera method. Details of these methods can be found in Appendix \ref{sec:beam_profiling}. %
Additionally, for Advanced LIGO and Advanced Virgo, calibrated cameras are placed in the near and far field of pick-off beams to track temporal changes in beam shape. For the core interferometer, these beams first pass through beam reducing telescopes which may introduce systematic errors. %
Depending on the accumulated Gouy phase, deviations in this beam size can tell the GWD scientists, operators and commissioners either about changes in $w_0$ or $z_0$. A similar approach is used for the Advanced LIGO input optics and further information can be found in~\cite{T1300941}.

For the input and output optics, single-digit-percent-level mismatching is achieved via traditional methods during commissioning. Once single-digit-percent-level mismatching is achieved, resonant methods must be used which directly interrogate the overlap of the incoming light and resonator eigenmode.

\subsubsection{Resonant wavefront sensing}
\begin{figure}
    \centering
    \includegraphics[width=\linewidth]{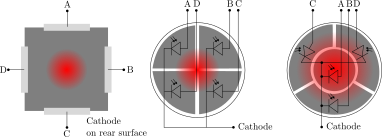}
    \caption{Frequently used segmented and position sensing diodes. Active areas are shown in dark grey, metal contacts in light grey and the beam is shown in red. Far left depicts a biased lateral effect position sensor, the middle depicts a quadrant photodiode and the far right depicts a bulls-eye photodetector.}
    \label{fig:segmented_diodes}
\end{figure}
In the case of alignment sensing, the beat between reflected PDH sidebands and reflected \HG{01}/\HG{10} modes is routinely used to control the alignment of suspended cavities (\!\cite{Morrison1994b,Fritschel98} and therein). Since the \HG{} modes are orthonormal, on a large area photodetector there would be no beat. However, a Quadrant PhotoDetector (QPD) breaks the orthonormality. Some segmented photodetectors are shown in \Figure~\ref{fig:segmented_diodes} and a historical review of the alignment sensing developments between 1984 and today is provided in Appendix~\ref{sec:alignment_sensing}. 
\begin{figure}
     \centering
     \begin{subfigure}[t]{\linewidth}
         \centering
         \includegraphics[width=0.75\linewidth]{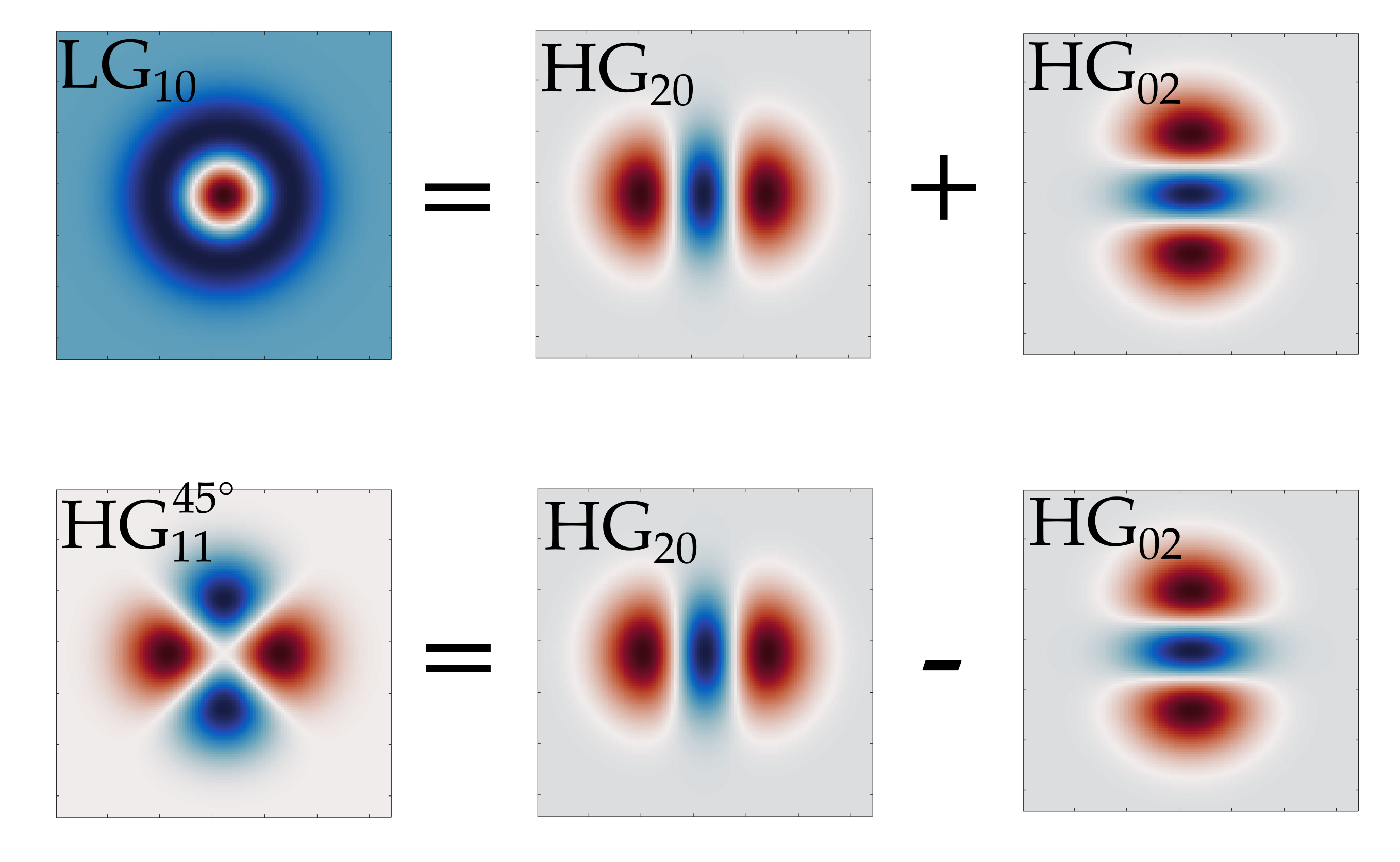}
         \caption{Decomposition of the $\LG{10}$ and 45\textdegree~rotated $\HG{11}$ in the HG basis.}
         \label{fig:modeconva}
     \end{subfigure}\\
     \begin{subfigure}[t]{\linewidth}
         \centering
         \includegraphics[width=0.8\linewidth]{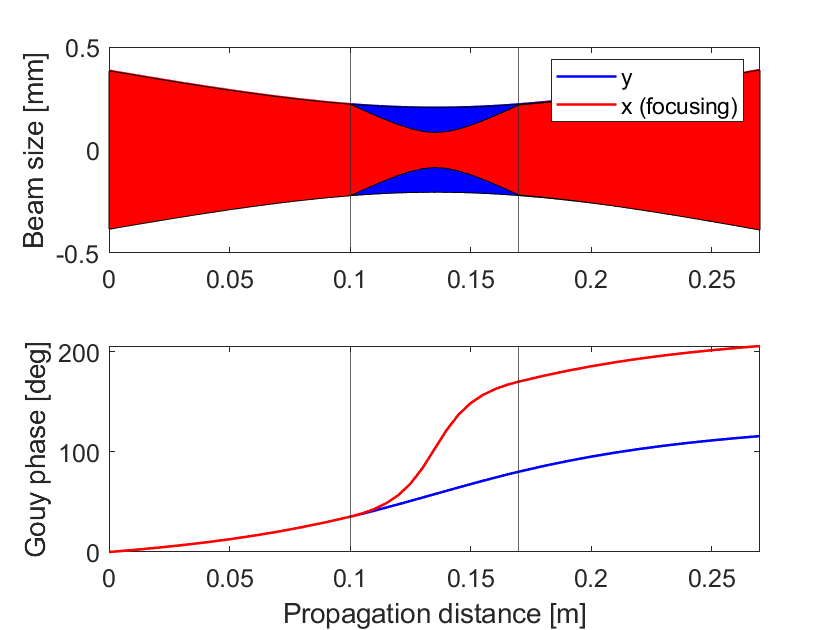}
         \caption{Plot of the beam size and Gouy phase accumulation on the x and y axis as the beam travels through a mode converter.}
         \label{fig:modeconvb}
     \end{subfigure}
        \caption{Mode conversion can be used to make the \LG{10} mode detectable on a QPD.}
        \label{fig:modeconv}
\end{figure}

In the year 2000, Mueller and others proposed extending the resonant wavefront sensing used for alignment control, to mode matching control~\cite{mueller2000}. For a small mismatch in beam parameters, light will scatter from mode \HG{n,m} into \HG{n+2,m} and \HG{n,m+2} with amplitude coupling coefficient~\cite{bayer-helms, mueller2000, ciobanu2020},
\begin{align}
    k_{n,n+2} &= \frac{i\sqrt{(n+1)(n+2)}}{4}\left(\frac{\Delta z - \Delta z_R}{\overline{z_R}}\right) \label{eq:knn}
    \\ \therefore \quad 
    k_{0,2} &= \frac{i}{2\sqrt{2}}\left(\frac{\Delta z - \Delta z_R}{\overline{z_R}}\right),
\end{align}
where $\Delta z$ denotes the difference in waist position, $\Delta z_R$ the difference in Rayleigh range, $\overline{z_R}$ the mean of the two Rayleigh ranges and $n,m$ are mode indices. If the mode mismatch is reasonably an-astigmatic then scattering will occur equally into \HG{02} and \HG{20} modes. The result is equal to a doughnut shaped \LG{10} mode~\cite{Beijersbergen93,Visser04,T1600189}, as illustrated \Figure~\ref{fig:modeconva}. Therefore, a bulls-eye photo-detector (BPD), shown schematically in \Figure~\ref{fig:segmented_diodes} breaks the orthonormality. Thus a beat occurs between the \LG{10} mode and the reflected \HG{00} cavity locking sidebands. By using two BPDs, it is possible to simultaneously gather information on $\Delta z$ \& $\Delta z_R$, provided the accumulated Gouy phase between the cavity and the two BPDs is 180 \& 270 degrees, respectively. Whilst promising, the technique has not been widely accepted by the community. BPDs are difficult to source, furthermore, overlapping requirements on the accumulated Gouy phase \& beam radius at the BPD mean that, likely, a custom BPD is required. 
\subsubsection{Mode conversion}
One option is to sidestep the BPD requirement by converting the \LG{10} mode associated with mode mismatch into a $\HG{11}$ \textit{pringle} mode which is more convenient for detection with a QPD. Because both HG and LG modes form a complete basis to describe laser beam modes, one can decompose $\LG{10}$ into a combination of two HG modes: \HG{20} and \HG{02}~\cite{Beijersbergen93,ONeil00,Visser04}. Additionally, a 45~\textdegree~rotated $\HG{11}$ mode can also be decomposed in the same combination of HG modes, minus a sign flip, as shown in \Figure~\ref{fig:modeconva}. So, in principle, by adding a $\pi/2$ phase shift along one axis, it is possible to convert $\LG{10}$ into a 45~\textdegree~rotated $\HG{11}$. In order to do this, the Gouy phase shift is exploited.

Generally, this is done using two cylindrical lenses to create a confined region where the beam is astigmatic. Provided the lenses are positioned so that the beam is no longer astigmatic outside of the converter, the additional Gouy phase shift along the astigmatic axis remains constant after the converter and subsequent detection is possible on a QPD. Consider a beam (along the non-focusing axis) with Rayleigh range, $z_R$ and beam-waist position, $z_0$. The desired Gouy phase shift will be achieved with cylindrical focal length~\cite{Beijersbergen93},
\begin{equation}
    f = \frac{z_R}{1+1/\sqrt{2}}.
\end{equation}
This lens must be placed at $z^\pm = z_0 \pm f\sqrt{2}$ to mode match the output beam while achieving the Gouy phase shift. \Figure~\ref{fig:modeconvb} shows the beam size in the astigmatic and non-astigmatic axis, and the evolution of the Gouy phase shift for a mode converter using cylindrical lenses. Mode converters with the same principle have also been realised using spherical mirrors at a 45\,\textdegree\xspace angle~\cite{Uren2019}.
\begin{figure}
    \centering
    \includegraphics[width=\linewidth]{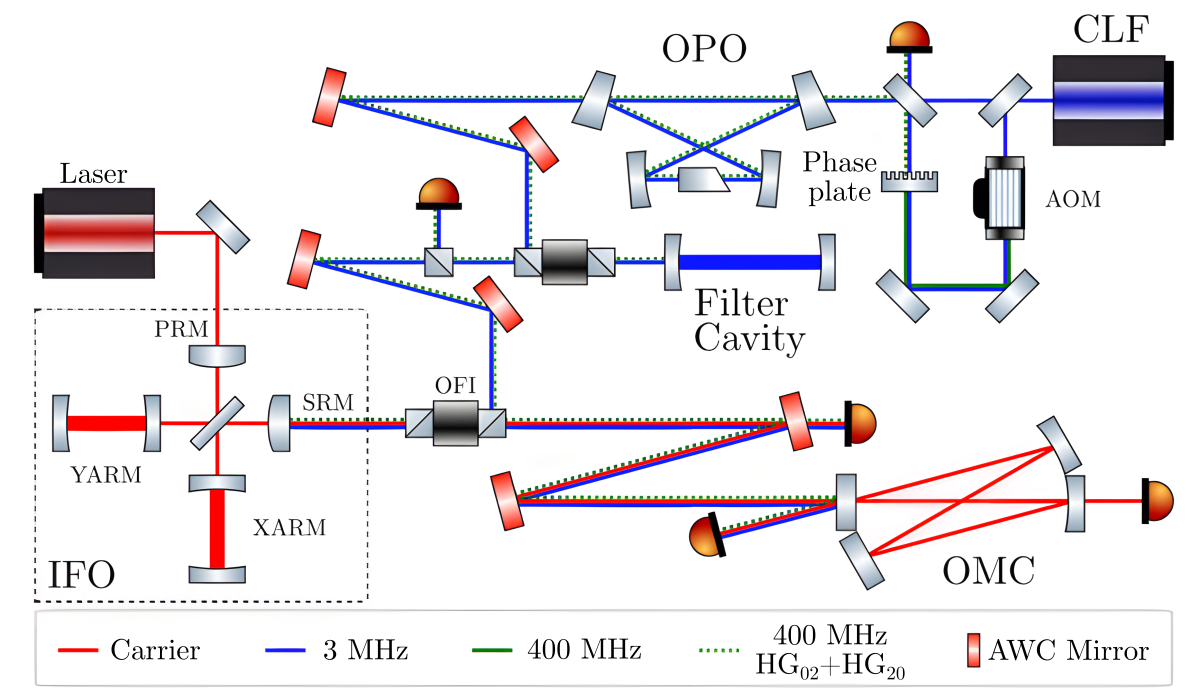}
    \caption{Suggested implementation of a RF Beam Shape modulation in a gravitational wave detector. The modulation is generated with a AOM and phaseplate, it then co-propagates with the squeezed light until the OMC. At the OMC it is reflected and the mode-matching can then be inferred. Reprinted with permission from~\cite{ciobanu2020}, \textcopyright\, The Optical Society. }
    \label{fig:rf_beam_shape_modulation}
\end{figure}

Typically, mode converters are used in conjunction with heterodyne techniques to generate an error signal for mode mismatch. In~\cite{Sandoval2019,MaganaSandoval18}, frequency-shifted sidebands in the $\HG{00}$ mode beat against the converted $\HG{11}$ on a QPD to generate an error signal. By placing a beam splitter and two sets of QPDs separated by 45 degrees in Gouy phase, it is possible to get separate error signals for waist size and waist position mismatch respectively~\cite{Sandoval2019}. Such a mode converter set-up is currently being used to mode match the filter cavity for frequency-dependent squeezing at Advanced Virgo~\cite{GrimaldiThesis}

One limitation of the mode converter is that it assumes a \LG{10} mode and is thus incompatible with astigmatism. The frequency dependent source uses curved mirrors in off-axis telescopes in order to reduce optical losses~\cite{Hello96}, causing some small astigmatism by design. This astigmatism limits how much mode matching can be achieved with the current mode converter~\cite{GrimaldiThesis}.

\subsubsection{Radio frequency beam modulation}
\label{sec:rf_beam_modulation}
Mode conversion is under active investigation, and is a promising technique for future gravitational wave detectors. However, it suffers from three key limitations. Firstly, it is not trivially generalized to sensing the coupled cavity mismatch. Secondly, it requires Gouy phase telescopes. Finally, it requires several new sensors and pick-offs to be introduced into the vacuum envelope. In 2017, reference was first made to the RF beam shape modulation method in an alignment sensing paper~\cite{Fulda_17}. In 2020, the first RF Beam modulation proposal paper was published~\cite{ciobanu2020}. The premise is to generate a frequency offset beam-shape modulation using an Acousto-Optic Modulator (AOM) and a phase-plate. Using the nomenclature defined in section \ref{sec:quantum_enhanced_ifo}, the carrier is at $\hgket{0,0}{q_1}{\omega_0}$ and the modulation is at $\hgket{2,0}{q_1}{\omega_0+\Omega} + \hgket{0,2}{q_1}{\omega_0+\Omega}$. Considering only the $n$ modes, where the beam experiences a basis change, we obtain
\begin{align}
    \hgket{0}{q_1}{\omega_0} \rightarrow &\phantom + k_{0,0}\hgket{0}{q_2}{\omega_0} + k_{0,2}\hgket{2}{q_2}{\omega_0} + \mathcal{O}(\Delta z^2, \Delta z_R^2)\\
    \hgket{2}{q_1}{\omega_0+\Omega} \rightarrow &\phantom + k_{2,2}\hgket{2}{q_2}{\omega_0+\Omega}\nonumber
        \\&+k_{2,0}\hgket{0}{q_2}{\omega_0+\Omega} \nonumber
        \\&+ k_{2,4}\hgket{4}{q_2}{\omega_0+\Omega} + \mathcal{O}(\Delta z^2, \Delta z_R^2),
\end{align}
where $k = k(\Delta z, \Delta z_R)$ denotes a coupling coefficient, defined in \Equation~\ref{eq:knn}. Further values of $k$ can be found in~\cite{ciobanu2020}. On a photodiode where the beam radius is much smaller than the active area, only modes of the same order will produce beat notes due to the orthogonality of the HG modes. Therefore, in the presence of a mismatch, there will be a beat with frequency $\Omega$ and the two mode-matching quadratures $z$ and $z_R$ can be read out with appropriate choices of demodulation phase. 

The scheme as proposed for LIGO is shown in \Figure~\ref{fig:rf_beam_shape_modulation}. The $\hgket{2,0}{q_1}{\omega_0+\Omega} + \hgket{0,2}{q_1}{\omega_0+\Omega}$ mode is generated from the Coherent Locking Field (CLF)~\cite{Vahlbruch06,Chelkowski07,Ganapathy22}, which passes through the Optical Parametric Oscillator (OPO) that generates the squeezing. The mode sensing field then co-propagates with the squeezing until it is reflected by the OMC. This proposal only requires only a single new component---the phaseplate, since all other components exist.

In contrast to generating the modulation with an AOM and phaseplate, several authors have begun exploring the possibility of using an electro-optic lens. First mentioned in~\cite{Fulda_17}, initial results were shown by two independent groups in 2020~\cite{G2001575,G2001619}. Proof-of-principal work was presented in 2023~\cite{Chiarini23} and the applicability of the schemes to higher-order carrier modes is discussed in~\cite{Tao23}.

Another approach is to forego the phaseplate and inject only the frequency offset beam. Thus relying on the mode mismatch to excite the higher-order mode. For example, consider the field $\hgket{0,0}{q_1}{\omega_0+\Omega}$, when the beam experiences a basis change, we obtain,
\begin{align}
    \hgket{0}{q_1}{\omega_0+\Omega} \rightarrow &\phantom + k_{0,0}\hgket{0}{q_2}{\omega_0+\Omega}\nonumber
        \\&+ k_{0,2}\hgket{2}{q_2}{\omega_0+\Omega} + \mathcal{O}(\Delta z^2, \Delta z_R^2).
\end{align}
On transmission of the cavity, the auxiliary modulation is filtered away leaving only $\hgket{0}{q_2}{\omega_0}$ and $\hgket{2}{q_2}{\omega_0+\Omega}$. In general, these do not produce a beat as they are different modes, however on a finite area photodiode or on a BPD a beat note will be produced. The scheme is an extension of the 1984 Anderson alignment sensing proposal~\cite{anderson84} and was recently published~\cite{Goodwin-Jones23}. Furthermore, by observing on transmission, there is the possibility of developing a scheme directly sensitive to the mismatch between cavities~\cite{Goodwin-Jones23}. However, the mode separation frequency of the second cavity must be within the FWHM of the first cavity. The $\hgket{0}{q_1}{\omega_0 + \Omega}$ sideband is then resonant in the first cavity. On a mismatch between the cavities it scatters into the $\hgket{2}{q_2}{\omega_0 + \Omega}$ mode with complex amplitude given by \Equation~\ref{eq:knn}. This technique is powerful as it is the first direct measure of mode mismatch between coupled resonators. Some proof-of-principal work has been carried out at LIGO Livingston~\cite{llo_alog_60763}.

\subsection{Beam Decomposition}
\label{sec:beam_decomp}
The previous section deals entirely with matching the complex beam parameter of an incoming light field to the complex beam parameter of a resonator. However, mirror surface roughness~\cite{Pinard2017}, parametric instability~\cite{Biscans19} and thermal transients can all degrade the mode basis~\cite{Ciobanu2021}. These degradations shift the mode basis away from the ideal HG one, which is only valid for infinite diameter, perfect, spherical mirrors in free space. 

In this abstract space, two figures of merit are important. Firstly, the overlap between the input laser eigenmode and the resonator eigenmode which dictates the power in cavity and, thus, the shot noise level of the detector. Secondly, the overlap between the squeezer eigenmode and the resonator eigenmode. This overlap dictates the maximum possible quantum enhancement. These two things must be optimised without causing resonances that lead to instabilities.

In the remainder of this section, we will review two key ideas. The first is the phase camera, a Cartesian decomposition of the wavefront, referenced to some characteristic beam. The second is to decompose the beam into some known reference basis, using either an optical convolution or reference cavity. 

\subsubsection{Phase Cameras}
\begin{figure*}
    \centering
    \includegraphics[width=0.8\textwidth]{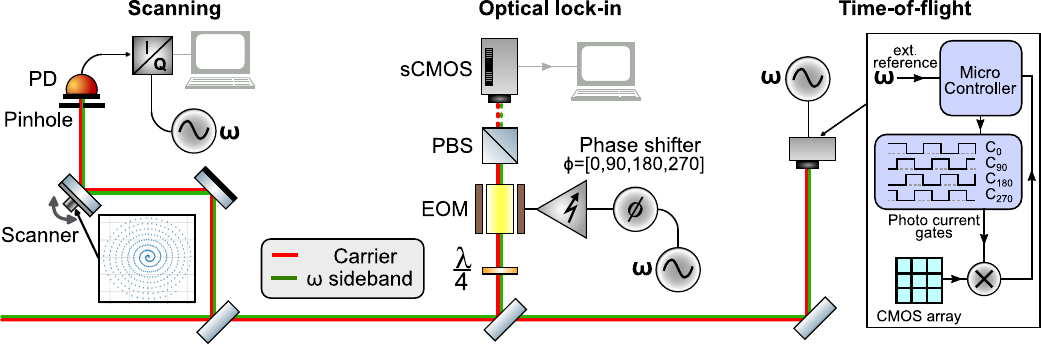}
    \caption{Schematics of the scanning, optical lock-in and time-of-flight phase camera designs. Not shown is the external reference beam which is required when individually measuring the sideband field(s), rather than measuring the beat of the carrier and sideband field which is as shown here.}
    \label{fig:phasecamcomparison}
\end{figure*}
Phase camera is the name colloquially given to devices that perform radio-frequency differential wavefront sensing at significantly higher spatial resolution than can be achieved with segmented photodiodes. The namesake relates to their ability to produce image maps of the transverse amplitude and phase of optical beats between frequency-shifted beams. By contrast, Hartmann wavefront sensors measure only the combined wavefront and segmented PDs can usually measure only one mode. 

There exist principally five designs of phase cameras which differ in their approach to measuring the spatial profile of the optical heterodyne field. The first, and most natural design would be a high element number PD array, where each element is simultaneously demodulated at the desired frequency. However, the typical frequency offset for a GWD side-band is 1-100\,MHz and so cross-talk between the elements poses challenging limits on the bandwidth and performance of these devices.
The three most developed designs of phase camera are highlighted in \Figure~\ref{fig:phasecamcomparison}, and the following paragraphs discuss each of these individually.

\paragraph{Scanning pinhole phase cameras}
\begin{table*}[hbt]
    \centering
    \begin{tabular}{l | p{2.2cm} p{3cm} p{2.2cm}}
      \hline \hline
         & scanning & optical lock-in & time-of-flight  \\
        \hline \hline
        Pixels (px) & 128x128 & 2048x2048 & 320x240 \\
        \hline
        Frame rate (fps) & 1 (max. 10) & 10 (max. 100) & max. 60\\
        \hline
        Sensitivity (dBc/px) & -61 (at 1 fps) & -62 (at 0.5 fps) \newline -72 (120x128~px, 1 fps) & -62 (at 1 fps) \newline -50 (at 7 fps)\\
        \hline
        Spot size change (\%) & 2.3 & 0.15 & 1.1 \\
        \hline
        Spatial precision (mode weight ppm) & 16500 & 1100 &  7800\\
        \hline
        Phase RMSE (nm) & 0.7 & Not available & 0.1\\
        \hline
        Maximum frequency & 250 MHz & 100 MHz & 100 MHz \\
        \hline
        Num. demodulations & 11 & 1 & 1 \\
        \hline \hline
    \end{tabular}
    \caption{Phase camera parameter comparison. The values presented were extracted or derived from~\cite{AgatsumaPC, schaaf16, cao2019, Muniz_Ballmer21} respectively. The spot size change is given by 3/(Num. Pixels) and assumes the spot diameter is 1/3 of the camera aperture. Spatial precision is computed from the spot size change using \Equation~\ref{eq:knn}. The sensitivity for the scanning phase camera was computed taking into account the optimal power in the reference beam~\cite{schaaf16}.} %
    \label{tab:phasecameras}
\end{table*}
The first phase camera was developed to image the differential wavefront of the GWD sidebands compared to the carrier~\cite{BetzwieserThesis}. In the first demonstration~\cite{Goda2004}, a frequency offset reference beam was combined with the test beam on a beamsplitter. The field was then reflected from a pair of galvanometer-driven steering mirrors onto a pinhole photodiode. By demodulating the photodiode signal at the difference between the reference field and the test fields, phase and amplitude maps could be obtained of the carrier and sidebands independently.

In the first applications of the phase camera~\cite{BetzwieserLIGOnote, BetzwieserThesis}, the reference field was no longer used. Instead, by demodulating at the difference between the sideband and carrier frequency, phase and amplitude maps could be obtained of the sideband referanced to the carrier wavefront. The technology was very mature and has been used to study the mode structure of initial LIGO~\cite{modeStructLIGO} and assist with commissioning the initial LIGO OMC~\cite{BetzwieserLIGOnote, BetzwieserThesis}.

The scanning phase camera has seen significant development since the first proposal. Modern cameras simultaneously image 11 sidebands (upper and lower sideband at 5 demodulation frequencies plus the carrier). Each image has $2^{14}$ pixels, with maximum phase resolution $\sim \lambda/1600$ (at the centre of the beam) acquired in under a second~\cite{AgatsumaPC}. In contrast to the first version, the cameras can scan either the test beam, or both beams over the pinhole. They make use of high-dynamic range ADCs, fast FPGAs and calibrated actuators to achieve this resolution. Due to the slow scanning speed when compared to environmental phase fluctuations, this resolution is only achievable for differential phase images between carrier and sidebands. See~\cite{AgatsumaPC} and references therein for further details. This style of phase camera is routinely used in the Virgo GWD~\cite{schaaf16}. There, differential phase images are preferentially obtained to mitigate phase fluctuations in the reference beam fiber and also the residual motion of the benches themselves.

\paragraph{Optical lock-in phase camera}
Optical lock-in phase cameras use amplitude modulation to optically demodulate the RF beat down to $\sim 100$\,Hz where it can be measured with a standard CCD camera.
The amplitude modulation is achieved with a Pockels cell \& waveplates, which first modulate the polarisation, and a Polarising Beam Splitter (PBS) which converts this into amplitude modulation.
The Pockels cell is driven at the desired sideband frequency, with large voltages $\sim1$~kV being required to achieve adequate modulation depths.
Amplitude \& phase maps are acquired by subsequently stepping the phase of the amplitude modulation through $\phi=[0,\pi/2,\pi,3\pi/2]$ and recording images.
The four images taken are then digitally processed into amplitude \& phase maps.

The technique was proposed in~\cite{cao2019}, where a 2\,Mpx camera is used to produce images with sensitivity up to -62\,dBc after 2\,s of averaging. Subsequently, the camera has been trialled at LIGO~\cite{G2001604} and used to image parametric instabilities~\cite{Schiworski2022}, thus determining the mode of the PI. Further work has explored using neural networks to determine the mode decomposition of the beam~\cite{Schiworski:21}. However, the camera can only image one sideband at a time. Future integration work will need to consider stable and efficient generation of the large driving voltage ($\sim 1\,$kV) at tune-able frequencies, with minimal RF contamination of the laboratory environment.

\paragraph{Time-of-flight phase camera}
Time-of-flight (ToF) cameras~\citeg{HansardBook} are a mature technology used in many augmented reality products, including smartphones and video games. The operating principal is that a set of infrared light emitters turn on at $t=T_0$ and off at $t=T_1$, and the cycle repeats at $T_2$. Four quadrants of a pixel are triggered to collect charges, at 4 different phases between the $T_0$ and $T_2$. By taking the ratio of the charges collected in each pixel, the distance can be estimated. A distance $L$ can only be measured unambiguously provided as long as it does not exceed the wavelength of the modulation, $L < c(T_2-T_1)/2$.

ToF cameras were proposed as phase camera by the Advanced Virgo team (section 7.7.1.3 in ref.~\cite{AdVTDR}). The light is discarded and the quadrants of each pixel are clocked at the frequency difference between the sideband and the carrier, similar to the optical demodulation approach. This means that only one spectral component can be interrogated at a time. Up to 100\,MHz demodulation with -62\,dBc has been demonstrated~\cite{Muniz_Ballmer21}. 

\paragraph{Spatial light modulator based phase cameras}
Spatial light modulator based phase cameras tag each pixel on a reference beam with an orthogonal code~\cite{Ralph17}. The code is imprinted on the reference beam with a spatial light modulator (SLM). This reference beam is combined with a test beam on a beamsplitter, which can be focused onto a single photodiode. By demodulating photodiode data, with the code for a particular pixel, information is obtained on the phase and amplitude of the beat note for that pixel. For an SLM with 9 pixels, 9 concurrent demodulations must take place. Further work is required to adapt the scheme to image the sidebands in a terrestrial GWD. 
\FloatBarrier
\paragraph{Summary}
Table \ref{tab:phasecameras} compares the performance of the current most developed types of phase cameras. For further details on using phase cameras to measure mode mismatch, see~\cite{ricardo_cabrita_2022_6758929}.
\subsubsection{Modal weighting}
\label{sec:mode_weighting}
Since both the HG and LG modes are orthonormal and form a complete basis, any electric field can be described by
\begin{align}
    E(\vec{r},t) = \sum_{n,m,j} a_{n,m,j} u_{n,m} \exp(i\omega_jt),
\end{align}
where $u_{n,m}$ denotes either the HG or LG modes. The process of determining the parameter $a_{n,m,j}$ for a given electric field is called modal weighting. One of the options to determine the modal weights is to perform a mode scan. To perform the mode scan, a special optical resonator is assembled which should be stable, with clear separation between different modes. Usually, a cavity like the LIGO pre-mode cleaner is used~\cite{UeharaSPIE}. Kwee et al. developed a diagnostic breadboard based on the pre-mode cleaner, that was able to perform the mode weighting along with measurement of the beam pointing and relative intensity noise~\cite{KSWD07}. This method was successfully used in 2015 to characterize thermal lensing in various materials~\cite{Bogan15} and has inspired follow-on work in the FSOC community~\cite{Mah19}. However, it is only possible to obtain the mode power, $|a_{n,m,j}|^2$ rather than the complex mode weight. 

In 2010, Takeno et al. developed an extension to the technique, where the phase and amplitude of deviations from a perfect \Gaussian mode could be imaged. The result is a phase camera that processes only deviations from the carrier mode~\cite{Takeno11}. 

\begin{figure}
    \centering
    \includegraphics[width=\linewidth]{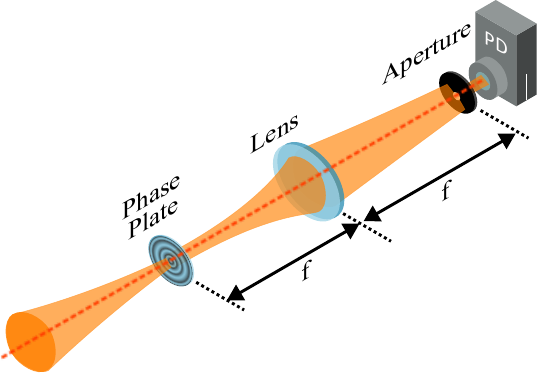}
    \caption{Mode weighting with a spatial filter. An incoming electric field $E(\xi,\eta)$ is passed through a phase modulating plate, with transmission function $T(\xi,\eta)$. In the far field the on-axis intensity is proportional to the overlap integral of $T$ and $E$. }
    \label{fig:mode_weighting}
\end{figure}

An alternative approach is to use a spatial filter to compute the fraction of light in a particular mode. This is achieved by passing the beam through a basic optical convolution processor (BOCP), consisting of a phaseplate, lens and aperture, all separated by the lens focal length, as illustrated in \Figure~\ref{fig:mode_weighting}. It is possible to show, by application of the Rayleigh-Sommerfeld equation, that the on-axis intensity is equal to
\begin{align}
    I(0,0,z_\text{PD}) \approx& \frac{\exp(i(2kf + \pi/2))}{f\lambda}\nonumber\\&\int\int_{-\infty}^\infty E(\xi,\eta,z_\text{Phaseplate})T(\xi,\eta) \mathrm{d}\xi\mathrm{d}\eta.
\end{align}
Such a derivation may be found in many places (e.g. \S 3.1~\cite{PhD.Jones}). If $T(\xi,\eta) = u_{\overline{n},\overline{m}}^*(\xi,\eta)$, then since the HG and LG modes are orthonormal, the on-axis intensity is proportional to the power in the $\overline{n},\overline{m}$ mode,
\begin{align}
    I(0,0,z_\text{PD}) \propto a^2_{\overline{n},\overline{m}}/f^2\lambda^2.
\end{align}
$f$ denotes the focal length of the lens and we have implicitly assumed the light is monochromatic~\citeg{Jones20} . This idea was first proposed by Golub in 1982~\cite{Golub82}. 

However, difficulty fabricating the photographic plates led to a slow adoption of the technique~\cite{Forbes16}. The popularity of liquid crystal based spatial phase modulators~\cite{Boruah09} has led to a resurgence in the technology~\cite{Flamm12}. Several authors have investigated methods of encoding amplitude information using a phase-only device. One popular method is to modulate the depth of a blazed grating~\cite{Davis1999} and exact solutions are possible~\cite{Bolduc13}. Another approach is to use \textit{computer generated hologram correlation filters}~\cite{Kaiser09}. The technique is now quite popular and there exists a dedicated review on the topic~\cite{Forbes16} along with an earlier text~\cite{GolubBook}, to which the reader is referred for further general information. 

\begin{table*}
\centering
\begin{tabular}{l|p{3.5cm}|l|l}
\hline\hline
                      & Precision [Mode Weight]   & Integration & Tested in GWD \\ \hline\hline
Spatial filter        & $0.6\,\text{ppm}/\sqrt{\text{Hz}}$~\cite{Jones21}\newline 4000 ppm (RMS) \cite{Jones20a}
                        & To LIGO CDS \cite{Jones21} & No \\ \hline
Diagnostic breadboard & 20\,ppm (RMS) \cite{KSWD07}                                                                                          & Complete  & All \\ \hline
Phase cameras & 1000\,ppm (1\,Hz) [Table~\ref{tab:phasecameras}]                                  & Complete  & LIGO, Virgo \\ \hline
Resonant wavefront sensing & $15000$\,ppm \cite{mueller2000}                                                                           &  Complete & LIGO  \\ \hline
Radio frequency beam modulation & $100$\,ppm (RMS) \cite{ciobanu2020}                                    &  No & No  \\ \hline\hline
\end{tabular}
\caption{Compairson of some of the direct sensing and mode decomposition technologies. Conversion from fractional waist radii to mode weight is obtained }
\label{tab:sensing_technologies}
\end{table*}
Within the gravitational-wave community, mode weightings at the part-per-million level are required. A thorough tolerance analysis found that the technique is often limited by the relative positioning accuracy and finite aperture effects, which cause cross-coupling of unwanted modes~\cite{Jones20}. Overcoming these limitations resulted in a mode (power) weighting precision of 4000\,ppm. Further investigation used a meta-material to achieve a mode (power) weighting precision of $0.6\,\text{ppm}/\sqrt{\text{Hz}}$~\cite{Jones21}, with an optical apparatus fully integrated into the LIGO Control and Data System. Further work is required on back-scatter and longitudinal tolerance analysis before this work could be integrated with a GWD. 

\subsection{Summary of Sensing Technologies}
In this section, we have reviewed several different schemes for sensing eigenmode mismatch. GWDs use a careful design procedure informed by the simulation to ensure their optical telescopes will achieve high magnification while also ensuring insensitivity to small radii of curvature errors. This is complemented by Hartmann cameras which image the surface deformations of the main optics. Several technologies have been proposed and trialed for direct mode sensing, including Resonant Wavefront Sensing, Mode Conversion, and Radio Frequency Beam Modulation. Additionally, several technologies have been proposed for beam decomposition including: Phase Cameras, Diagnostic Breadboards, and Spatial Filters. We present a comparison in Table~\ref{tab:sensing_technologies}.

\FloatBarrier
\section{Precision actuation of spatial quantum states}
\label{sec:actuators}
The required actuation on mirror radii of curvatures in advanced GWDs ranges from $\sim 100$ \textmu D on the test masses~\cite{Brooks16} to $\sim 100$ mD on the input and output optics~\cite{T1900144}. Many adaptive optics technologies already exist, but are unsuitable for use in a GWD. Requirements on scattered light exclude many mirror technologies. For example, spatial light modulators are excluded by wide angle scatter caused by the pixel grid. It is challenging to polish ultra-thin mirrors to meet the scatter requirements and this excludes many unimorph~\citeg{Alaluf18}, monomorph~\citeg{Raphae17} and various bimorph~\citeg{ALCOCK201387, Kudryashov15} mirrors. Mirrors must be suspended from complex seismic isolation chains to suppress phase noise and thermal noise and this excludes many mechanical and piezo-actuated mirrors due to vibrations, phase-flicker and $1/f$ noise~\citeg{Ralph17, Wlodarczyk14, Samarkin16}. Finally, vacuum outgassing requirements exclude many commercial solutions. 

The gravitational-wave community has coalesced around three sets of solutions. Firstly, for the core interferometer, transmissive optics are radiatively heated to introduce thermal lensing. Secondly, the input path must survive high optical powers and so transmissive (or reflective with the HR surface after the substrate) optics are heated by direct thermal contact. Lastly, the output path has ultra-low-loss and phase noise requirements, so reflective optics, which are mechanically stressed, are used.

\subsection{Core interferometer}
Within the core interferometer, we will review four approaches used to develop appropriate radiative heating patterns on the optic. 
\subsubsection{Ring heaters}
\label{sec:ring_heaters}
\begin{figure}
  \begin{subfigure}[b]{0.3\linewidth}
    \centering
    \includegraphics[trim={8cm 0 0cm 0},clip,width=\linewidth]{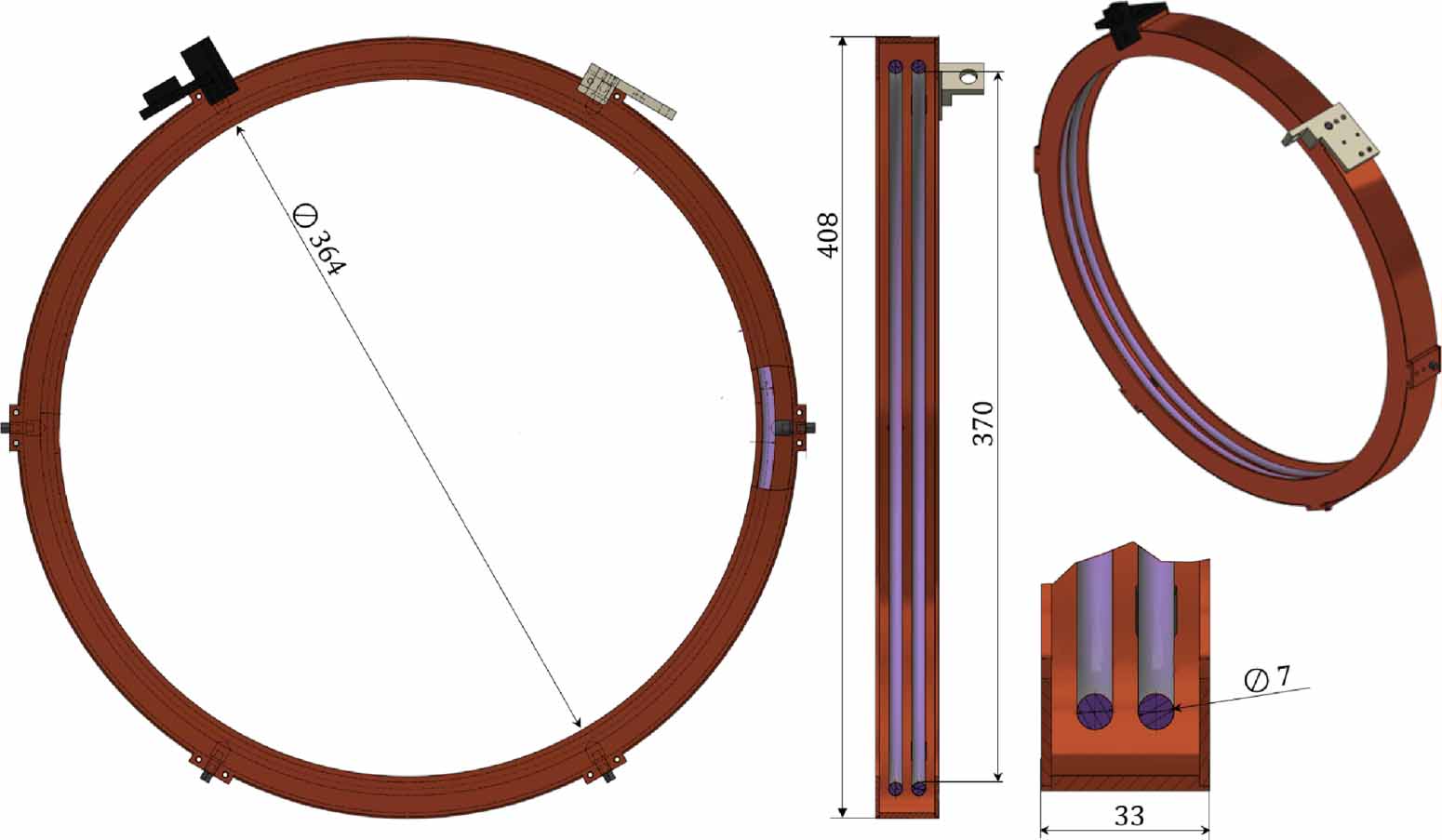}
    \caption{Drawings for the ring heater in Advanced Virgo}
    \label{subfig:rh1}
  \end{subfigure}\hfill
  \begin{subfigure}[b]{0.35\linewidth}
    \centering
    \includegraphics[width=\linewidth]{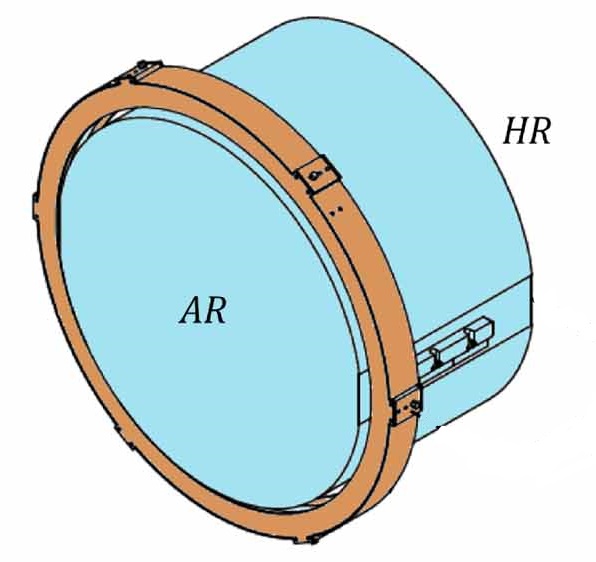}
    \caption{Depiction of ring heater around test mass}
    \label{subfig:rh2}
  \end{subfigure}\hfill
  \begin{subfigure}[b]{0.25\linewidth}
    \centering
    \includegraphics[trim={1cm 0 1cm 0},clip,width=\linewidth]{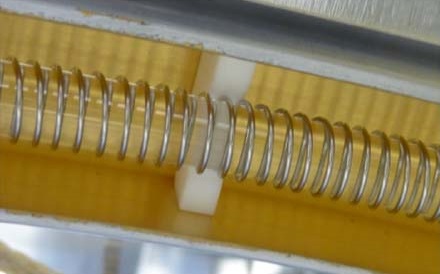}
    \caption{Photograph of a Ring Heater installed in Advanced LIGO.}
    \label{subfig:rh3}
  \end{subfigure}
  \caption{Ring heaters. (a) and (b) are reproduced from \cite{Nardecchia2023} under the CC-BY-4.0 Licence. (c) is reproduced with permission from~\cite{Brooks16}, \textcopyright\, The Optical Society.}
  \label{fig:RH}
\end{figure}

Ring heaters were discussed as early as 2002~\cite{Lawrence02} and first tested at the GEO600 observatory~\cite{Lueck04}. These ring heaters can be used to compensate for thermal lensing and also to correct for small radii of curvature errors. While thermal lensing mainly occurs at the centre of a mirror, because that is where the laser beam deposits its energy, the ring heater heats the outside barrel of a mirror. \Figure~\ref{fig:RH} illustrates the concept. 

Fused silica has both a positive refractive index change with temperature, $\mathrm{d}n/\mathrm{d}T>0$ and a positive coefficient of thermal expansion, $\mathrm{d}L/\mathrm{d}T>0$. For the ITM, the recycling cavity eigenmodes are affected by both $\mathrm{d}n/\mathrm{d}T$ and $\mathrm{d}L/\mathrm{d}T$. The arm cavity eigenmode is only affected by $\mathrm{d}L/\mathrm{d}T$ on both test masses. Therefore the heating at the edge creates a more convex lens, offsetting the central heating caused by absorption of the \Gaussian laser beam. The heaters are made of two pyrex rings surrounded by a polished copper shield, which reflects the radiation onto the test mass. The pyrex ring is heated from a conductive wire wrapped around~\cite{Brooks16,Aiello2019,Nardecchia2023}.

Because the ring heater acts by reducing the thermal gradient, it can only induce a reduction of the radius of curvature of a mirror. For a typical fused silica mirror with a nominal radius of curvature around 1.5~km, the ring heater at maximum power is able to reduce the radius of curvature by 100~m ($\sim 100 $\textmu D) \cite{Nardecchia2023,Brooks16}.

A variant of the GEO600 ring heater is also used on the signal recycling mirror 3 (SR3) in Advanced LIGO. In this instance, a ring behind the test mass causes a thermal expansion of the outer edges of the mirror, thus making SR3 more convex~\cite{G1501373}.

In Advanced Virgo, ring heaters have been installed also to tune the radii of curvature of the recycling mirrors (both PR and SR) and of the optics the filter cavity for the injection of the squeezing~\cite{Nardecchia2023}.

Recently, Richardson et al. have investigated extending the use of the ring heater concept to the front surface of the test mass. Referred to as a FROnt Surface Type Irradiator (FROSTI), it would allow a non-spherical radius of curvature to be created by thermally tuning a spherical optic~\cite{G2201439}. Thus changing the resonance condition of very high order optical modes, as discussed in section~2\ref{sec:design_considerations}.

\subsubsection{Central heating radius of curvature correction}

Peripheral heating solutions like ring heaters can only reduce the radius of curvature. While this allows to compensate for thermal effects (thermo-optic and thermo-elastic), however, there may be situations where it is necessary to increase the radius of curvature of an optic, for example, to compensate for deviations from nominal specifications. The central heating radius of curvature correction (CHRoCC) projects a heat pattern onto the center of an optic, increasing the radius of curvature. The device is a black body emitter inside a heat shield, together with an ellipsoidal reflector used to project the heat into the target optic~\cite{AccadiaCHRoCC2013}.

It was initially proposed and developed to correct the radius of curvature of the end test masses in enhanced Virgo. The change of the radius of curvature of the mirrors shifts the resonance frequency of higher order modes, which were otherwise degenerate with the fundamental mode, causing locking instability issues~\cite{AccadiaCHRoCC2013}. In this instance, the dynamic range was almost 1000~m change in the radius of curvature. Currently, these CHRoCC devices have been installed to increase the tuning possibilities of the power and signal recycling mirrors in Advanced Virgo.

\subsubsection{CO\textsubscript{2} lasers and compensation plates}
\label{sec:CPs}

Shortly after the GEO600 demonstration of a ring heater~\cite{Lueck04}, compensation plates were proposed~\cite{Lawrence02, Degallaix04} and tested at the High Optical Power Test Facility, Gingin, Australia~\cite{Zhao06}. The premise is to install a dedicated optic, with strong optical path length change with temperature. One concept, was to use materials where either $\mathrm{d}n/\mathrm{d}T <0 $ or $\mathrm{d}L/\mathrm{d}T < 0$ (\cite{Mueller2002} and therein). This has been successfully implemented for Faraday Isolators in GWDs~\cite{Palashov2012,Mueller16}. However, as Zhao explains \cite{Zhao_PC} it was difficult to find suitable materials for use in a cavity. 

The interferometer is much less sensitive to noise on the CP than noise on the test mass, and so actuation can be much stronger~\cite{Degallaix2005}. For example, in the first proposal, a conductive ring heater was glued onto the compensation plate which would not be possible on the test mass. However, in later revisions, this was swapped for a powerful CO$\mathrm{_2}$ laser~\cite{Blair_PC}. The CO$\mathrm{_2}$ emits at a wavelength of 10.6$\mathrm{\mu m}$, which is strongly absorbed by fused silica. The absorbed light changes the optical path length through the material and tuning the intensity of the laser adjusts the change to optical path length~\cite{Aiello2019, Brooks16}. In aLIGO, the CP was initially polished flat~\cite{E1000158} and then re-polished with some focal length offset~\cite{E1400394}.

The CO$\mathrm{_2}$ laser and CP can be used as either a positive or negative lens~\cite{Aiello2019, Brooks16}. When shining a \Gaussian beam at the center of the compensation plate, the temperature rises in the middle and the radius of curvature of the optic is increased. This mode of operation is called central heating.

In Advanced Virgo, the aberrations due to the average coating absorption, which are mostly radially symmetric, are corrected for by shining two annuli-shaped ring patterns on the compensation plates. The optimal compensation heating pattern was estimated using a FEA simulation~\cite{AdVTDR}. These two ring-shaped intensity patterns are obtained with two axicon lenses. A combination of waveplates and lenses is used to control the intensity and thickness of each ring respectively. This set-up is called the Double Axicon System (DAS)~\cite{Aiello2019}.

Finally the CO$\mathrm{_2}$ laser can also be used in combination with a scanning system~\cite{Lawrence02, MudaduThermalSS}, to compensate for non-axi-symmetric aberrations. The beam is moved with a constant speed over the compensation plate and the intensity is adjusted for each location according to a correction map. Additionally, work is ongoing at Virgo~\cite{VIR-0373A-23} to use the modified Gerchberg-Saxton algorithm~\cite{Mehrabkhani17} and a commercially available deformable mirror to shape the CO$\mathrm{_2}$ laser beam and develop novel heating pattern beams.

\subsubsection{Matrix heaters}

Matrix heaters are arrays of heaters that can illuminate a heating pattern onto the test masses. This heating pattern deforms the optic surface, leading to the enhancement or suppression of particular optical modes. The first proposal~\cite{Day2013}, was to use 9 $\times$ 1\,cm $\times$ 1\,cm pixels operating between 500 and 1200\,\textdegree C. The proposal suggested placing this outside the vacuum system and imaging it onto the test mass using a ZnSe lens. The heating pattern could then be used to tune problematic higher-order-mode resonances in the arm cavities. One of the study's coauthors then developed a thorough optimization procedure, that could be used to minimize wide-angle scatter caused by mirror surface roughness~\cite{Vajente2014}. 

The first demonstration of a matrix heater~\cite{Wittel18} was at GEO600. %
198 individually addressable heating elements were arranged on a PCB. The heating elements could dissipate 1\,W at around 600\,\textdegree C ($\lambda \gtrsim 4 $\,\textmu m). In this instance, the imaging apparatus consisting of a ZnSe vacuum window is used along with a parabolic aluminum mirror. 

The matrix heater concept has been further developed at Virgo~\cite{point_absorber_virgo_gwadw2023,point_absorber_virgo_grass22}, to cope with localized heating from highly absorbing regions of the test mass, with a characteristic size of about $100$\textmu m, known as point absorbers~\cite{Brooks21}. These absorbers cause scattering of power from the fundamental mode to higher order modes in the arm cavities, thus increasing the round-trip-losses. The peculiarity of the Virgo solution is that the corrective pattern is produced through the use of a binary mask illuminated by a single heating element. This allows to greatly simplify the driving electronics and to increase the spatial resolution of the actuator, being equivalent to a 40 $\times$ 40 array of heaters. The mask is then imaged onto the test mass using a germanium lens and ZnSe vacuum window.

\subsection{Input path}
\label{sec:act_input}
The idea to place some smaller active optical elements in the vacuum system naturally arises following the development of the compensation plates. The initial design for aLIGO had provision for two adaptive optical elements~\cite{T1300941}. The design was similar to the initial compensation plate. In this instance, a circular aperture SF57 substrate was used as a lens~\cite{G0900115}. This lens was held in place by four segmented metal clamps in thermal contact with the barrel of the optic. The clamps were heated by a resistive wire wrapped around the clamp. The substrate was polished to be flat on both sides, by heating the clamps the optical path length on the edge of the mirror could be increased, thus, a negative lens could be formed with dynamic range $f \ge 10$\,m~\cite{G1200866}. The transfer function was stable, with a unity gain frequency of around 20\,mHz, DC gain of around 100, and it did not significantly degrade the beam quality~\cite{G1100359}. By actuating each quadrant independently, tilt and astigmatism could be introduced into the output~\cite{G0900115}. However, the optic was never installed as the mode matching in the input optics was never considered a performance limitation~\cite{Fula_PC}. Furthermore, there were concerns about excessive heat on the table interfering with the seismic isolation~\cite{G1200866}. 200\,mD of actuation was achieved with 10\,W of heating~\cite{Arain10}. The work is summarised in~\cite{Arain10}

The idea can be expanded by using a mirror and placing the AR surface at the front and the HR surface at the back. In this way, the beam passes through the substrate twice. On the HR surface, a multielement heating array can be bonded. An initial proposal used 9 heating elements and found minimal hysteresis and excellent linearlity~\cite{Canuel12}. 64\,mD of focal length actuation was achieved with only 160\,mW, reducing the thermal load on the isolated tables. Meanwhile promising results were shown for astigmatism actuation. In the 2013 proposal~\cite{Kasprzack2013}, 61 heaters were used, enabling the correction of Zernike modes up to fifth order. 

\subsection{Output and squeezing path}
\label{sec:output_and_squeeze}
The output and squeezing paths of current GWDs require $\sim 90\%$~\cite{o4_sqz_ligo} mode matching to ensure excellent quantum efficiency between the cavity used to generate the squeezing, filter cavity, interferometer and output mode cleaner~\citeg{mcculler20, toyra17}. This path has similar requirements to the input path. However, instead of high power handling, requirements exist on backscatter, phase noise and loss. As justified in~\cite{Cao20}, these requirements justify a new class of solutions such as thermally actuated bimorph mirrors, with high dynamic range. The operating principle is that a 6\,mm thick mirror is inset into an aluminum ring. The inner diameter of the aluminum ring is smaller than the outer diameter of the mirror, and thus, applies a compression bias. The compression bais crushes the mirror to be more convex. As the mirror is heated, the compression bias is decreased. Further work is ongoing to demonstrate concave mirrors and larger diameters~\cite{GoodwinJones24}.

The excellent noise characteristics of the thermally actuated mirror make it ideal for use in the GWD output paths. However, for the filter cavity paths, it was required to dither the radii of curvature at $\mathcal{O}$(1-10\,Hz). Since the unity gain frequency of the thermally actuated mirrors is $\mathcal{O}$ 1\,mHz, a new technology was required for the filter cavity path. The adopted solution is a carefully engineered piezoelectrically actuated mirror. The design uses preloading and a custom flexure to convert the stress from the piezo into a spherical deformation while meeting the aLIGO noise requirements~\cite{Srivastava22}. 

\section{Applied mismatch mitigation strategies}
\label{sec:sites}
A mathematical derivation of dynamics of thermal issues in gravitational-wave test masses was published in 2009~\cite{vinet09}. Summaries of the thermal compensation systems in the core interferometer of Advanced Virgo were published in 2011~\cite{Rocchi11}, 2019~\cite{Aiello2019} and 2023~\cite{Nardecchia2023}. A summary of the thermal compensation system in the core interferometer of Advanced LIGO was published in 2016~\cite{Brooks16}. In this article, we will briefly summarise the sensing schemes and actuation channels that are routinely used throughout the entire interferometer in both LIGO and Virgo. 

\subsection{The input optics}
The active optics for the input optics is a critical subsystem, as higher order modes can mediate the transfer of noise into the GWD readout. For alignment, resonant wavefront sensing (see appendix~\ref{sec:alignment_sensing}) is routinely used. Actuation is then provided by coil-magnet actuators (\cite{Cooper23} and therein).

For sensing the beam waist position and radius diagnostic, calibrated cameras are placed in the near and far field of the beam~\cite{T1300941}. These cameras are placed between the IMC and PRC, to track the eigenmode in the IMC. This eigenmode will change as IMC heats up. In addition, pick-off photodiodes in the input path and PRC monitor the power recycling gain. The power recycling gain depends on mode matching efficiency amongst other things. 

Section~3\ref{sec:CPs} discussed the idea of optics with a negative optical path length change with respect to temperature. This was not used for compensation plates, however, the Faraday isolators in the GWD input optics do contain a thermally driven negative lens~\cite{Palashov2012}. %
\subsection{Squeezed light injection and interferometer output path}
Active optics on the squeezed light and interferometer output paths are critical to realizing high levels of quantum noise suppression. All the techniques specified in the previous section are used. 

There are various options to sense the mode matching between the interferometer and OMC. One commonly used approach is to partially lock the interferometer and use the OMC as a diagnostic breadboard (section 3\ref{sec:mode_weighting}). However, there are several issues. For example, the output beam contains many sidebands and junk light, which contaminates the measurement of mode matching. Additionally, the length sensing and control scheme for the full interferometer requires the OMC to be locked, therefore the state of the interferometer during the OMC scan does not represent the state used for gravitational wave detection. 

Another useful measure is the complex transfer function between the test masses motion and displacement readout~\cite{llo_alog_60763}. Maximizing the gain of this transfer function indicates high optical power in the arm cavity and, thus, good mode matching. However, without linear error signals, this cannot be automated. Furthermore, the function depends on many interdependent parameters and cannot be uniquely constrained to mode matching. 

For the squeezed path, the squeezing level into the interferometer and the audio diagnostic field~\cite{Ganapathy22} are used to infer the squeezing mode matching. In terms of actuators, both thermally and piezoelectric actuated bimorph mirrors are used (section 3\ref{sec:output_and_squeeze}). However, as with the transfer function method, it cannot trivially be automated and requires substantial commissioning time to diagnose issues. 

\subsection{The core interferometer}
\begin{figure*}
    \centering
    \includegraphics[width=.8\linewidth]{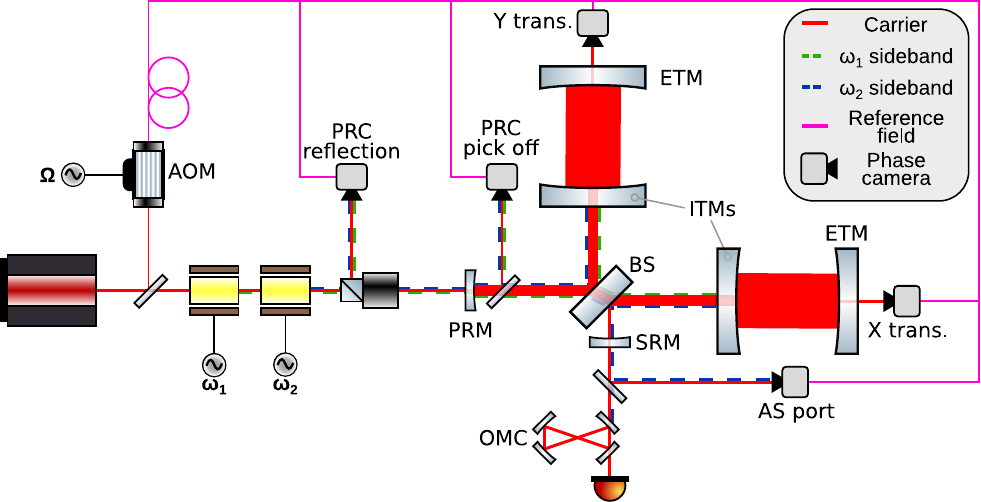}
    \caption{Possible phase camera locations and schematic diagram of phase cameras in a GWD. In Advanced Virgo, cameras are installed at the PRC pick-off and AS Port.}
    \label{fig:phasecamlocations}
\end{figure*}
Within the core interferometer, it is exceptionally challenging to define a mode basis. Despite efforts to produce perfect spherical mirrors, effects such as point absorbers~\cite{Brooks21}, apertures and surface figure errors shift the mode basis away from the HG basis. This leads to a shift in the accumulated Gouy phase. The problem is further complicated by thermal transients which add a time-dependent Gouy phase change. The Gouy phase change can bring unwanted modes onto resonance, destroying the sensing and control schemes. 

At the time of writing, current detectors use a variety of tools to infer to the desired operating point. The complex transfer method described above is frequently used. In addition, maximizing the power recycling gain and scans of the arm cavities are also used to infer the state. However, at LIGO the only dedicated mode sensing actuator is the Hartmann sensor. The Hartmann sensor is only sensitive to changes in the thermal state and does not offer a quantitative decomposition of the mode content in the arms. Thus, it is not possible to clearly identify higher-order mode resonances throughout the core interferometer, nor it is trivial to extract mode-sensing error signals for automatic mode matching. Actuation is provided by: the ring heaters, compensation plates and SR3 heater. 

At Virgo, the use of marginally stable recycling cavities amplifies the sensitivity to mode mismatch. In addition to the methods above, Virgo makes routine use of scanning pinhole phase cameras. Two of these devices are currently installed, the first one at the output port of the interferometer and the second one on a pick-off beam extracted in the PRC. \Figure~\ref{fig:phasecamlocations} shows a schematic diagram of the phase cameras in a GWD. Another phase camera is planned in reflection of the PRC in the future. In the Virgo implementation, a frequency-shifted reference beam is obtained from a pick-off in the main laser beam. The pick-off is fibre-coupled to the phase camera installation. The shifted reference beam is useful for differentiating the upper and lower sidebands, which is useful as sideband imbalance is known to cause issues in the locking of cavities~\cite{KeikoRAM}. However, phase noise along the path of the optical fiber, during the measurement time ($\sim 1$\,s) limits the accuracy of the phase measurement allowing, nowadays, only differential phase measurements between carrier and sidebands. Some optical fiber phase noise cancellation techniques are under test to be implemented in future upgrades in order to be able to independently measure the absolute phase of the different beams (carrier, upper and lower sidebands). Moreover, the information provided by the images acquired by the phase cameras have been used to estimate the mode content of the carrier beam at the output of the interferometer~\cite{SchaafThesis}, giving inputs on the actuation needed to reduce the resonance of higher-order modes in the FP cavities. An optimization of the data processing of the phase camera images is in progress with the aim of estimating the mode content of all the acquired beams. This will help in the evaluation of the mismatch level in the various cavities of the GWDs.
\section{Future requirements}
\label{sec:the_future}
Two next-generation GW detectors are anticipated to be operational around 2035: Cosmic Explorer (CE) in the US~\cite{Reitze2019} and Einstein Telescope (ET) in Europe~\cite{design_study_update_et}. In addition, we anticipate the routine use of integrated quantum photonics to produce vast entangled networks~\cite{Simon2017}, imaging below the shot noise level~\citeg{Casacio2021} and define SI units~\cite{Lisdat2016}. In this section we will begin by reviewing the challenges faced by future GWDs, we will then discuss the work required to integrate the mode sensing solutions described in section~\ref{sec:sensors} into these detectors. Lastly, we will close with a discussion on how this work could be applied outside the field of GWD. 

\subsection{Future requirements for gravitational-wave detectors}
ET will be an underground facility, hosting 3 pairs of interferometers, each 10\,km long with 200\,kg test masses. Each pair will contain a low-frequency (ET-LF) and high-frequency detector (ET-HF). The reference design for ET-LF is a cryogenic detector with mirrors at 10\,K and only 18\,kW of optical power in the arms. The reference design for ET-HF is to operate at room temperature with 3\,MW of optical power in the arms. 

CE is a conceptual design for a 40\,km long detector with 1.4\,MW of optical power and 320\,kg test masses. The reference design for CE is a singular, broadband interferometer with the potential to build a 20\,km second interferometer, possibly in the southern hemisphere~\cite{ce_horizon_study}. There exists scope for a potential upgrade to CE, referred to as CE2, which may integrate longer wavelength lasers and cryogenic silicon optics. Both CE and ET can operate independently or as part of a global network, enabling unprecedented cosmological reach and high detection rates~\cite{ce_horizon_study}.

Several intermediate detectors and prototypes have been proposed to bridge the technological gap between the current and future detectors. Current detectors have around 40\,kg, room temperature test masses with around 300\,kW of optical power. One intermediate detector is the Neutron Star Extreme Matter Observatory (NEMO) which uses 4.5 MW of stored optical power and 70\,kg mirrors at around 130\,K mirrors~\cite{nemo}. Another is the LIGO Voyager project which proposes 3\,MW of optical power with 50\,kg, 123\,K mirrors~\cite{Adhikari_2020}. The cryogenic community is supported by several prototypes~\cite{ETPathfinder, ETEST_CDR, mariner_gwadw, gingin_gwadw}.

The Virgo project is developing Advanced Virgo Plus, Phase II~\cite{Acernese2023} which uses 100\,kg room temperature mirrors and will operate in the 5th Observing Run. The LIGO community is currently using the term A\# to refer to the upgrade in the LIGO vacuum enclosure, following the 5th observing run. This will likely involve heavier test masses, 10\,dB of squeezing and 1.5MW of stored optical power in the arms~\cite{T2200287}.  The Virgo community is using the term Virgo\_nEXT for a similar upgrade~\cite{VIR-0617A-22}. The larger test masses may suppress many radiation-pressure-mediated angular instabilities (\cite{Liu_and_Bossilkov18} and references therein), therefore enabling higher power operation. Furthermore, larger test masses may permit larger beam sizes, reducing coating thermal noise~\cite{vinet09}. 

\subsection{Semi-classical mode sensing and control challenges in future detectors}
These changes required by future detectors will have several important effects on the interferometer modes. 

First, the free spectral range of the longitudinal resonance of arm cavities is close to the detection band, around 3.8\,kHz for CE and $\sim$15\,kHz for ET. This means that the majority of higher-order modes are within the detection bandwidth. Even whilst operating on the fundamental mode, Brownian motion on the test mass scatters light into higher-order modes. As such, when within the detection bandwidth, their control and suppression is significantly more difficult.

When a high overlap is achieved between a mirror mechanical mode and an optical mode, the optical mode can become resonant at a frequency offset. The radiation pressure may then drive the mechanical mode, causing the system to become unstable. The effect is referred to as Parametric Instability (PI, e.g.~\cite{Evans15} and therein), and the parametric gain depends on the optical power and accumulated Gouy phase, amongst other things. As the stored power increases, the effects of PI will become more severe~\cite{Pan2022}. Furthermore, as higher-order modes are brought into the detection band PI will be harder to control. As such it will be critically important to minimize scatter into higher-order modes and avoid PI amplification.

Future detectors will likely use larger diameter test masses. These large dimensions will put strict requirements on the polishing of the optical elements. Surface figure errors over the large area of the test mass contribute to shifts away from the HG mode basis, thus making an analytic description of the interferometer state intractable.

High powers in the interferometer will lead to large transient thermal effects in mirror substrates, especially in the central beamsplitter. This is especially true for ET-HF and CE, which use fused silica and operate at room temperature. In the current generation of detectors, this time-dependent mismatch has caused numerous problems. For example, by changing the mode shape and basis in the recycling cavities, the resonance conditions for the control sidebands in the first-order HG modes are shifted. Since these sidebands are used for angular sensing and control, changing resonance conditions introduces additional noise into the detector. Additionally, a changing mode basis means the PI gain becomes time-dependent and substantially complicates PI mitigation. Lastly, differential changes in mode basis couple laser frequency noise into the interferometer output signal, further limiting sensitivity. Given these constraints, the current detectors are able to operate using $\sim$1\,W of ring heater power \cite{llo_alog_karla_1W_RH_65817} and limiting to $\sim$300\,kW arm power~\citeg{capote_gwadw,llo_alog_valery_320kW_64807,llo_alog_vlad_320kW_64243}. %

A comprehensive study of the actuator requirements for further increases in LIGO arm power has been carried out~\cite{G2300624} and it was determined that increasing arm power will require increased ring heater and  CO\textsubscript{2} power as well as potential new actuators such as FROSTI. 
 
For cryogenic detectors, the thermal conductivity of crystalline silicon at $\sim$100\,K is $\sim$900\,W/m K~\cite{Glassbrenner64}. In contrast, the thermal conductivity of Fused Silica at room temperature is $\sim$1\,W/m K~\citeg{Combis2012, Wray2004}. %
When used for laser mirrors, the effective thermal conductivity of cryogenic silicon is slightly increased~\cite{Klemens1981}. A trade-off study on cryogenic suspensions for high-power GWDs found that geometric distortions will be suppressed by several orders of magnitude compared to their room temperature counterparts~\cite{Eichholz20}.

\subsection{Challenges on the way to 10\,dB squeezing}
Both ET and CE reference designs require 10\,dB of quantum noise suppression to reach design sensitivity.
In order to suppress both quantum shot noise at high frequencies and quantum radiation pressure noise at low frequencies, quantum uncertainty has to be squeezed in different quadratures at different signal frequencies.
Such frequency-dependent squeezing is achieved by reflecting squeezed light off a filter cavity~\cite{Kimble02,mcculler2020frequency,Zhao20,Acernese2023_FDS,o4_sqz_ligo}.
Reaching 10\,dB of squeezing requires the total optical loss to be less than 10\% in the full detection band. 
There are several sources of squeezing degradation, such as direct optical loss in the filter cavities and the detector, scattering loss, length and phase noises, coupling two quadratures of squeezed light and, importantly, mode mismatch~\citeg{Kwee2014, geo600_6db, Tse2019, Acernese2019}.
The overall effect of mode mismatch would have to remain at a level below $\sim$1\,\% (p.~120~\cite{design_study_update_et} and \S\,8.3.5~\cite{ce_horizon_study}).

Mode mismatch affects squeezing level in two distinct ways: as the direct source of optical loss, and as an additional degradation due to a dephasing mechanism\,\cite{Kwee2014, toyra17, mcculler20}.
In this mechanism, the squeezed vacuum coherently couples into a higher-order mode and then back into the signal mode after acquiring some (unknown) phase.
Due to this additional phase, the back-coupled light contributes a part of anti-squeezed quadrature into the squeezed mode, thereby degrading it.
The coupling phase depends on the specifics of the coupling mechanism, and the detector needs to be optimized with respect to it.

For CE and ET-HF, the large beam radius will require significant focusing, likely before and possibly also after the central beam splitter~\cite{Rowlinson21}. This focusing will incur a significant Gouy phase shift between different HOMs.
Various imperfections along the propagation path, most notably due to thermal effects on the central beam splitter, will introduce an additional unknown coupling phase between different HOMs.
Thus, contributing to the dephasing mechanism and leading to a significant reduction on squeezing beyond the direct loss.

ET-LF is planned to operate in a detuned regime, where the opto-mechanical interaction between the light and test masses causes an additional resonant signal enhancement for a range of signal frequencies, the resulting quantum noise level is thus rather complex. Filter cavities are required to replicate the quadrature rotation caused by the interferometer. As such, the filter cavity arrangement for ET-LF is commensurately complex requiring either two filter cavities (section~D.3 of~\cite{ET-0106C-10}) or one coupled filter cavity~\cite{PJones20}, to achieve the required quadrature rotation.
The dephasing mechanisms will cause significant degradation at the frequencies around the optomechanical resonance by coupling a large portion of anti-squeezing into the signal mode. 
This will require limiting the amount of squeezing injected into the detector, thus impacting the overall sensitivity.

\subsection{Recommendations}
We identify three clearly separable issues in future room temperature GWDs. Firstly, the coupling of optical states between resonators, especially in the recycling cavities, output optics and squeezing optics. Optimal mode matching cannot be achieved by indirect methods due to basis mismatches. For this reason, we recommend the use of one of the direct mismatch sensing schemes (section~2\ref{sec:direct_mm}), to be included in the baseline design of future interferometers. In addition to the direct correction, squeezed HOMs can be employed to further improve the sensitivity~\cite{toyra17, steinlechner2018mitigating}.

Secondly, the issue of the mode basis degradation in the core interferometer, caused by point absorbers, marginally stable cavities and higher-order effects. Without a clear understanding of the mode basis, it is not possible to understand the transient behaviour of the full interferometer. For this reason, we recommend the use of a dedicated beam-decomposition technology to be included in the baseline design for future detectors. Furthermore, we note that future detectors may need actuators able to make non-spherically symmetric wavefronts corrections, in order to mitigate higher-order effects. 

Lastly, the issue of the transient behaviour of the interferometer. Both Hartman sensors and other more novel approaches would be able to image the transient behaviour. However, additional work is required to develop new actuators to manage the transient thermal distortions in future detectors. 

For cryogenic detectors, further work is required to understand the requirements on actuation. Depending on the design of the recycling cavities, it may be possible to leverage the experience obtained from KAGRA as the detector moves to higher powers. For cryogenic detectors operating at 2000\,nm wavelength, neither silicon nor InGaAs photodetectors have suitable responsivity. Extended InGaAs photodetectors are widely available and so mode sensing solutions requiring photodetectors can be reworked for this wavelenght. However, technologies requiring quadrant or bulls-eye photodetectors, or cameras, may be prohibitively expensive, depending on wider market conditions.  

\subsection{Beyond Gravitational Wave Detectors}
\label{sec:beyond_gwd}
Photonic states have a broad range of applications. In this section, we will review a collection of applications and how the mode sensing technologies could be used to further these areas of research. 

One area with particular synergies is quantum information technology. 
The recent rise of photonic quantum computation~\cite{zhong2020quantum,madsen2022quantum}, where squeezed modes provide the resource for creating large-scale entanglement~\cite{takeda2019toward,fukui2022building}, requires extremely high levels of mode-matching and precise mode shape manipulation~\cite{kashiwazaki2023over}.
Building extended quantum networks with quantum computers (not only photonic) requires secure and efficient quantum state transport between different nodes. This is enabled by (continuous-variable) quantum key distribution (QKD)~\citeg{Pirandola2020} and quantum teleportation~\citeg{Simon2017}. Given the fixed loss per unit length of modern communication fibres, it is likely that such links will involve, at least, one free-space component to a satellite receiver~\citeg{Simon2017}. However, atmospheric turbulence adds a temporally fluctuating mode basis shift which may be corrected with the AO~\citeg{Wang2019}. This is similar to the temporal basis shift caused by thermal effects in the interferometer, which degrades the coherent and squeezed states in the interferometer. 
As such, these applications could benefit from the direct mismatch sensing schemes discussed in section~2\ref{sec:direct_mm}. Depending on the application, such research may also benefit from applying the ultra-low phase noise actuators discussed in section~3\ref{sec:act_input} and 3\ref{sec:output_and_squeeze} to avoid mixing quantum states. 

Several authors have considered the use of higher-order spatial modes to further enhance communication bandwidths~\cite{Richardson13,Wang18}. As shown in~\cite{Jones20a}, the fidelity with which the optical state must be matched depends strongly on the mode order. This area is analogous to matching a laser beam to an optical resonator and therefore many of the technologies discussed in section~\ref{sec:sensors} can be applied directly. 

Several quantum technologies have proposed using higher-order transverse modes to reduce their noise. For example, in the area of optical clocks, higher-order modes have been proposed to reduce thermal noise~\cite{Zeng18} or avoid point defects~\citeg{Jiao20}. Then, in the area of cavity assisted atomic interferometry~\cite{dovale17,Canuel18}, higher-order modes can be used to increase the size of the beam and, thus, capture more atoms. Finally, in the field of optical tweezers increasing higher-order mode indices corresponds to increasingly steep potentials~\cite{Mestre10}, leading to improved trapping. In all of these cases, switching to a higher-order transverse mode will lead to increased mode matching requirements~\cite{Jones20a}. In all three cases, matching to an optical resonator is required and as such many of the sensing and actuation schemes could be applied. In particular, the RF sensing schemes discussed in section~2\ref{sec:rf_beam_modulation} have commensurately stronger error signals~\cite{Tao23} and could be used to optimally match the beam into these cavities. This could reduce the linewidth of cavities in optical clock experiments, capture more atoms in atomic physics experiments and increase the strength of the potential in optical tweezer experiments. 

Quantum imaging technology can benefit from higher-order modes in several ways. First, entangled spatial modes allow higher resolution and lower noise in conventional imaging approaches~\cite{kolobov2000quantum,wagner2008entangling,treps2003quantum}. However, so far these applications were limited by the ability to create, control and detect quantum correlations in higher-order modes. Recent success in generating~\cite{treps2003quantum,zhang2022quadrature,li2022higher,Heinze2022observation} and observing~\cite{embrey2015observation,gabaldon2023quantum} of such states opens the way for their practical use, assuming they can be efficiently controlled. 
A second application in quantum imaging is the super-resolution approach, where higher-order modes are used to extract additional phase information and enable sub-diffraction imaging~\cite{taylor2013biological,tsang2016quantum,tsang2017subdiffraction}.
\section{Summary}
AO is one of the most important optical techniques with a wide variety of applications. Gravitational-wave physicists applied active optics as early as 1985 and applied AO as early as 2003. Following nearly three decades of development, AO has been successfully translated to the \Gaussian laser optics domain. With the introduction of frequency-dependent squeezing in observing run 4, we are seeing the routine use of AO to overcome quantum noise in a practical application. 

In this review, we grouped mode-sensing solutions into three categories: indirect mismatch sensing, direct mismatch sensing, and beam decomposition/basis identification. For each area, we have summarised the state-of-the-art mismatch sensing technologies and carried out a historical review of the technology to date. 

We have further grouped actuators into three categories by area of application. Firstly, the core interferometer, which requires handling $\sim 10^{6}$\,W of optical power, $\sim 100$\,\textmu D actuation ranges, $\sim 10$\,cm optics, and the phase noise less than $\sim 10^{-19}\,\text{m}/\sqrt{\text{Hz}}$. Secondly, the input path which must handle $\sim 100$\,mD actuation range and $\sim 100$\,W of optical power. Finally, the output path requires $\sim 1$\,W of optical power, $\sim 100$\,mD actuation range, and variable bandwidth and actuation noise requirements. For each category, we have presented several solutions which meet the requirements and indicated which ones are currently in use. 

The current detectors have employed a number of technologies to reach as much as 6\,dB of shot noise suppression without introducing additional QPRN and maintaining $\sim 300$\,kW of circulating power. In section~\ref{sec:sites}, we summarised the use of resonant wavefront sensing, diagnostic measurements, phase cameras, and actuators to achieve this result. 

Future gravitational wave detectors will require exquisite control of the spatial properties of the wavefront to avoid a plethora of issues. In section~\ref{sec:the_future}, we have summarised these requirements and discussed the range of proposed solutions that could be employed to resolve them. We closed with a discussion of how this technology could be translated into a plethora of domains from laser communications to quantum sensing and the authors look forward to a bright future of enhanced adaptive quantum optics. 
\begin{backmatter}
\bmsection{Funding}
This research was conducted with the support of the Australian Research Council Centre of Excellence for Gravitational Wave Discovery (CE170100004). D.D.B. is the recipient of an ARC Discovery Early Career Award (DE230101035) funded by the Australian Government.  Netherlands Organization for Scientific Research (NWO) (``First direct detection of gravitational waves with Advanced Virgo (VIRGO)'', project number 162).
M.K. was supported by the Deutsche Forschungsgemeinschaft (DFG) under Germany's Excellence Strategy EXC 2121 ``Quantum Universe''-390833306. 
R.C. is funded under the IISN convention 4.4501.19 “Virgo: physics with gravitation waves” of the Fonds National de Recherche Scientifique

\bmsection{Acknowledgements}
The authors are grateful to their LIGO, Virgo and KAGRA collaboration colleagues for helpful discussions and their contributions to the detectors. Specifically, we wish to thank our LIGO PnP and Virgo DRS reviewers. We thank Vaishali Adya, David Gozzard and Carl Blair for helpful discussions. This document has been assigned LIGO Document Number \href{https://dcc.ligo.org/P2300282}{P2300282} and Virgo Document Number \href{https://tds.virgo-gw.eu/ql/?c=19640}{VIR-0769A-23}. A.W.G. acknowledges the LIGO Scientific Collaboration Fellows program for supporting their research at LIGO Livingston Observatory.

\bmsection{Disclosures}
The authors declare no conflicts of interest.

\bmsection{Data availability}
No data were generated or analyzed in the presented research.

\bmsection{Supplemental document}
Supplemental documents follow as appendices.

\end{backmatter}
\bibliography{sample,aj_thesis_extracted}
\appendix
\clearpage
\FloatBarrier 
\section{Introduction to \Gaussian modes}
\label{app:formalism}
\begin{figure}
    \centering
    \includegraphics[width=\linewidth]{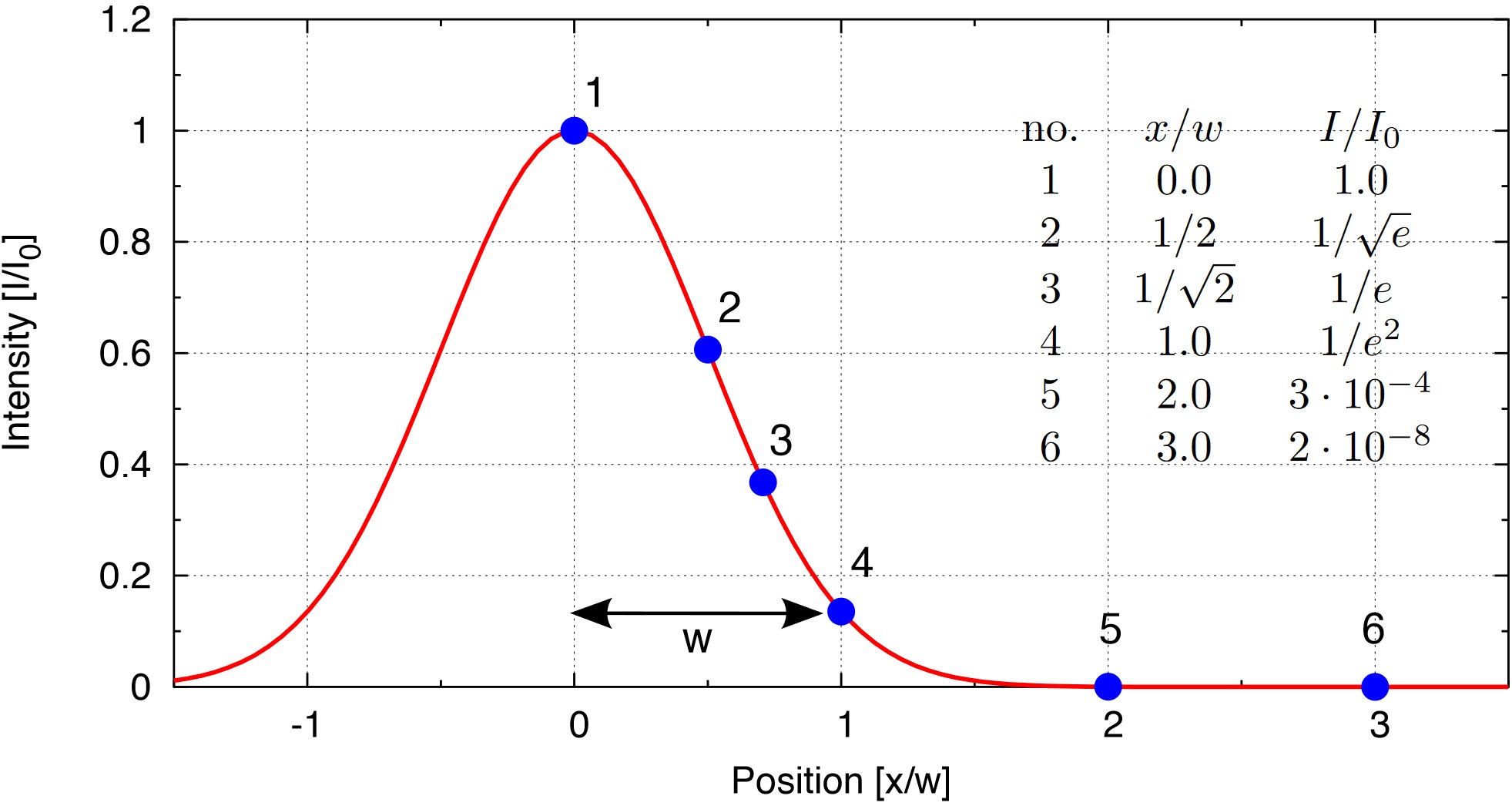}
    \caption{One dimensional cross-section of a \Gaussian beam. The width of the beam is given by the radius $w$ at which the intensity is $1/e^2$ of the maximum intensity. Reproduced with permission \cite{Bond2017}.}
    \label{fig:intensity_profile}
\end{figure}
The eigenmodes of optical resonators, formed with spherical mirrors are the \Gaussian modes. For total power, $P$, the intensity pattern of the lowest order mode is,
\begin{align}
    I(x,y,z) = \frac{2P}{\pi w^2(z)}\exp \left(\frac{-2(x^2+y^2)}{w^2(z)}\right)
    \label{eq:gaussian_beam}
\end{align}
where the optical axis is in the positive $z$ direction. $x$ and $y$ are Cartesian axes normal to both $z$ and each other. $w(z)$ is the beam radius at some point $z$ and acts as a scaling parameter. See figure~\ref{fig:intensity_profile} for an illustration. We can define the entire evolution of the beam by the minimal radius $w_0$ and the position of the minimal radius $z_0$, or combine both of these parameters into a single complex beam parameter
\begin{align}
    q(z) = i\frac{\pi w_0^2}{\lambda} + (z - z_0), \tagrepeat{eq:q}
\end{align}
where $\lambda$ is the wavelength. The first term has dimensions length and is indeed a characteristic length over which the beam varies in size, referred to as the Rayleigh range,
\begin{align}
    z_R \equiv \frac{\pi w_0^2}{\lambda}.
\end{align}
Each resonator with spherical optics defines a unique $q$, which defines the spatial eigenmode of the resonator. Putting this together, 
\begin{align}
    w(z) =w_0\sqrt{1 + \frac{(z-z_0)^2}{z_R^2}}.\label{eq:beam_radius}
\end{align}

In addition to the fundamental \Gaussian beam, a laser resonator can support a number of higher order modes. The first to show these higher order modes were Fox and Li~\cite{Fox61}. Over the next 5 years, Boyd, Gordon, Kogelnik, Goubau and others developed the theory of \Gaussian optics~(\!\cite{KogelnikandLi66} and references therein). Concepts such as resonator stability, HG modes, complex beam parameters and mode matching were derived. In 1986, Seigman published the seminal text \textit{Lasers}~\cite{Siegman}. A recent review containing a summary and physical interpretation relevant to GWD is \cite{Bond2017}. The HG mode with indices $n$ and $m$ is,
\begin{align}
    u_{nm}(x,y,z) &= u_n(x,z)u_m(y,z),
\end{align}
with
\begin{align}
    u_n(x,z) =& \left(\frac{2}{\pi}\right)^\frac{1}{4}\sqrt{\frac{\exp\left[i(2n+1)\Psi(z)\right]}{2^nn!w(z)}}H_n\left(\frac{\sqrt{2}x}{w(z)}\right)\nonumber\\&\exp\left(\frac{-ikx^2}{2R_C(z)} - \frac{x^2}{w^2(z)}\right),
\end{align}
where $R_C(z) = z + z_R^2/z$ denotes the Radius of Curvature and $\Psi(z) = \arctan (z/z_R)$ the Gouy phase. $H_n(x)$ are the (Physicist) Hermite Polynomials. Please see \S 9.4 of \cite{Bond2017} for further discussion and physical interpretation of these parameters. 

\subsection{Mode Mismatch}
Each optical resonator defines its own mode basis and pairs of lens or curved mirrors can be used to transform one mode basis into another. The usual formulation is the ABCD matrix approach (for example, see \S 9.13 of \cite{Bond2017}). The use of lens to match the complex beam parameter of the incoming light to the complex beam parameter of the cavity is referred to as mode matching. In 1984 Bayer-Helms described the mode matching problem  as decomposition of a \Gaussian beam into higher order modes \cite{bayer-helms}. The scattering between some mode \HG{nm} in basis $q_\text{in}$, into a resonator mode \HG{n'm'} in resonator basis $q_\text{cav}$ is given by the inner product of the modes. In general, the Bayer-Helms solution requires a numerical integration, however first order coupling coefficients have been derived for the case of mode mismatch and misalignment~\cite{ciobanu2020}. In the case of a translational misalignment $\delta$ and beam axis rotation $\gamma$ it is,
\begin{align}
    k_{n,n+1} \approx \left(\frac{\delta}{w_0} + \frac{q}{|q|}\frac{\pi w(z)\gamma}{\lambda}\right)\sqrt{n+1}.
    \label{eq:knn1}
\end{align}
In the case of a mode mismatch it is,
\begin{align}
    k_{n,n+2} \approx \frac{1}{4}\left(\frac{i\Delta z}{z_R} - \frac{\Delta z_R}{z_R}\right)\sqrt{(n+1)(n+2)}.
\end{align}
Mathmatically, this illustrates the well known experimental result that, for \HG{00} input, first-order modes in the cavity scan indicate alignment errors and second order modes indicate mode matching errors.

\section{Ground-based GWDs}
\label{sec:quantum_enhanced_ifo}

There are five, operational, audio-band, gravitational-wave detectors which form a single effective all-sky observatory. These are the LIGO Hanford and Livingston detectors~\cite{AdvancedLIGO15shortened} in USA, Virgo~\cite{AdvancedVirgo15shortened} in Italy, GEO600~\cite{Affeldt14,Dooley2016} in Germany and KAGRA~\cite{Akutsu2020, Aso13} in Japan. In addition, a sixth detector, LIGO India is planned~\cite{Padma19}. The 2015 first observation of a gravitational-wave signal~\cite{gw150914} was the result of decades of scientific endeavour coordinated across many continents. The science case for future detectors demands an instrument with extraordinarily differential length sensitivity ($\ll 10^{-20}\,\mRtHz$) between 5\,Hz and a few kHz. These audio-band GWDs are kilo-meter scale enhancements of the famous Michelson-Morley Interferometer~\cite{michelson1887}, where the passage of a gravitational wave causes a differential change in the arm length of the Michelson interferometer~\citeg{Sathya09}.
This change results in a relative phase change between the two beams, leading to the change in interference between them. Optical power leaks towards the output port and GWs can be detected~\cite{gw150914}.

Outside the audio-band, various technologies exist for GWD, including atomic interferometry~\cite{Graham2013,Canuel18}, space-based and pulsar-timing-based detectors~\cite{Bailes21}, lunar seismology~\cite{Harms2021} and levitated micro-disks~\cite{Aggarwal22}. In this work, \textit{GWD} refers to the audio-band Michelson-type detectors. %

All modern GWDs use high optical powers to enhance the signal. Current detectors achieve hundreds of kilowatts of laser power in the arms, and several megawatts are planned for future detectors.

To achieve this power buildup, the detectors employ two approaches. Firstly, they use Fabry-Perot resonant cavities as the arms of the Michelson interferometer. The cavity mirrors are Bragg reflectors coated onto $\sim40$\,kg input and end test masses (ITM and ETM). The test mass positions and laser frequency are tuned for resonance, leading to an increase of power between the cavity mirrors. 

A second approach to increase the circulating power is the addition of a so-called power recycling mirror (PRM) between the laser source and the beam splitter (BS). The working point of the interferometer is tuned to have a nearly perfect destructive interference for the light beams coming back from the arm cavities. Thus, most of the light is reflected back towards the laser source. However, by placing a mirror (PRM) between the BS and the main laser this light is recycled and injected back into the interferometer, providing an additional resonant enhancement of light power in the arms~\cite{meers88}. 

Further sensitivity enhancement is achieved by placing a signal recycling mirror (SRM) between the BS and the output port of the detector, amplifying the signal and opening the possibility to shape and optimise the detector bandwidth for specific GW sources (\!\cite{meers88,Heinzel96, Strain91, phd.Heinzel} \& references therein).

\begin{figure}
\centering
\includegraphics[width=\linewidth]{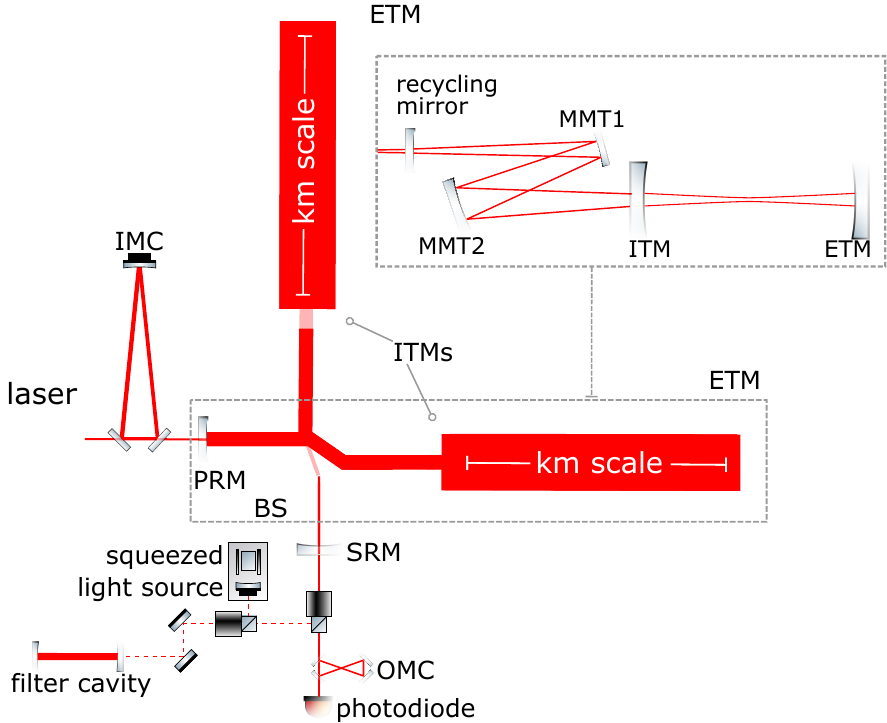}
\caption{Typical layout of an advanced gravitational-wave detector in a dual recycled Fabry-Perot Michelson interferometer configuration. Beam width is a qualitative measure for power. Inset shows the \Gaussian beam width in the coupled cavity including a mode matching telescope in the recycling cavity and ignoring the beam splitter. Extraneous optics are not shown and more details are found in text.}
\label{fig:GWdetector}
\end{figure}

The resulting layout is shown in \Figure~\ref{fig:GWdetector}. The phrase \textit{core interferometer}, is often used to describe the coupled cavity formed between the PRM, SRM, beam splitter and two kilometer-scale Fabry-Perot cavities. We use the term \textit{laser} to describe a series of lasing devices, pre-mode-cleaners and amplifiers. This light is further filtered through an input mode cleaner (IMC) cavity, resulting in a spectrally narrow and spatially pure \Gaussian beam. 

Various auxiliary radio-frequency (RF) modulations are used to control the interferometer. For example, Pound-Drever-Hall and derivatives are used for length sensing and control~\cite{drever83,Black01,Fritschel:01,Strain03,Allocca2020}. Various external modulations are also used for angular sensing and control and we provide a brief summary of those in Appendix~\ref{sec:alignment_sensing}. These modulations would cause noise on the GW readout photo-diodes, and so an Output Mode Cleaner (OMC) cavity is placed immediately before the photo-diodes to filter out RF modulations.

Throughout this review, we will make extensive use of the phrase \textit{input optics}, which refers to the optics before the PRM, while the phrase \textit{output optics} refers to the optics between the SRM and photodiode. The squeezed light source is generally included in the output optics.

Given the temporal and spatial purity of the beams in GWDs, we treat the beam as a monochromatic \HG{00} mode. Impurities and imperfections are described as frequency sidebands and higher-order HG modes and follow the mathematical formalism described in~\cite{Bond2017,Siegman}. In some cases we will make use of the compact notion $\hgstate{n,m}{q}{\omega}$, to describe an optical mode, with angular frequency $\omega$ in the \HG{n,m} mode with complex beam parameter $q$. For further justification, please see Appendix \ref{app:formalism}.

A comprehensive introduction to GWDs can be found in~\cite{VBP2,Saulson_Book_2nd,Grote2019,Blair2012}.

\section{Quantum-enhanced Gravitational Wave Detectors}
\label{sec:quantum_enhanced}
In the case of LIGO Livingston during the Observing Run 3, during which the detector achieved the lowest (published) noise~\cite{O3_instr_paper}, quantum noise limited the sensitivity at frequencies $\gtrapprox 20$\,Hz.  In this frequency range, the quantization of light leads to two distinct quantum noise phenomena. Below $\sim 40$\,Hz quantum fluctuations in amplitude quadrature of the light lead to random radiation pressure forces being applied to the test masses. This manifests as quantum radiation-pressure noise (QRPN) in the measurement record. Above $\sim 40$\,Hz, quantum fluctuations in the phase quadrature of the light lead to shot noise upon photodetection. Both of these quantum noise effects are fundamental in nature and arise from the Heisenberg uncertainty principle.
The sum of two noises is referred to as the Standard Quantum Limit (SQL)~\cite{Edelstein1978,Caves1980,Braginsky96b}, which is not possible to overcome without using quantum correlations.
For example, increasing optical power leads to improved shot noise at the expense of increased QRPN, which ultimately limits the use of high optical power in the detectors. The resulting quantum noise remains above the SQL. 

Another example of enhancing the shot-noise at high frequencies is the use of squeezed states of light, which allow a reduction in the uncertainty in one quadrature at the expense of another, maintaining Heisenberg uncertainty relation~\cite{Caves81, Schnabel2017}. A squeezed state, generated by a squeezed light source~\cite{Schnabel22_Review} is injected from the output port of the interferometer, between the OMC and SRM.
Since the first demonstration of quantum-enhanced detector GEO600~\cite{SqueezingNature11}, squeezed light became the main tool for operating the detectors at high sensitivity\,\cite{geo600_6db,Tse2019,Acernese2019}.
Following the success of GWD, squeezed light is nowadays utilised for dark matter searches\,\cite{backesQuantumEnhancedSearch2021}, applied quantum sensing\,\cite{Taylor2016,Li2018}, quantum communication\,\cite{gehring2015implementation,suleiman202240} and quantum computing\,\cite{fukui2022building}.

However, the straightforward application of squeezed light does not allow improvement in quantum noise in the full detection band, overcoming the SQL.
Reaching higher sensitivity requires the extensive use of quantum correlations\,\cite{Danilishin12,Danilishin2019}, such as frequency-dependent squeezing~\cite{Kimble02}, as currently employed by Virgo~\cite{Acernese2023_FDS} and LIGO~\cite{o4_sqz_ligo}.
This approach changes the phase of the squeezed quadrature for different frequencies by reflecting squeezed light off a specially designed filter cavity\cite{Kimble02}, as illustrated in \ref{fig:GWdetector}. 
The resulting sensitivity has both QRPN and shot noise suppressed at the same time and overcomes the SQL.

\begin{figure}
    \centering
    \includegraphics{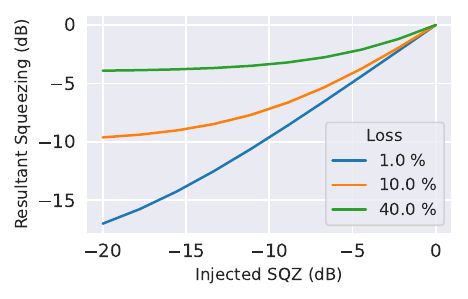}
    \caption{Achievable squeezing, for given squeezing injection and loss. Note that the loss fundamentally limits the achievable squeezing~\cite{Danilishin12,PhD.Schreiber}).}
    \label{fig:squeezing}
\end{figure}

Currently, detectors operate with around 6\,dB squeezing~\cite{geo600_6db}, despite the best-achieved table-top squeezing being as high as 15\,dB~\cite{Vahlbruch2016}. Further increase on squeezing level, required by future detectors, is limited by optical losses~\cite{Tse2019,Acernese2019}, which destroy the correlations between the sidebands of squeezed field. 
The maximum available squeezing is a function of the total loss in the squeezed light path. The optical losses act as an effective beamsplitter, coupling in unsqueezed vacuum state~\citeg{Danilishin12,PhD.Schreiber}. As shown in \Figure~\ref{fig:squeezing}, the losses in the system strongly dictate the maximum achievable squeezing level, according to
\begin{equation}
    S = S_0(1-l) + l,
\end{equation}
where $S_0$ denotes the ratio of squeezing to vacuum variance before the loss occurred, $S$ after the loss occurred, and $l$ denotes the fractional power loss in the system. By considering infinite initial squeezing, it is clear that the losses limit the achievable squeezing to $10\log_{10}(l)$\,dB. Phase noise further degrades this effect~\cite{Dooley2015}.%

Optical fabrication~\cite{Pinard2017} limits the ideal mode matching in advanced GWDs, causing losses and limiting power build-up. Furthermore, a small fraction of the light in the interferometer is absorbed by the mirror substrates and coatings. This absorption causes the mirrors to deform~\cite{vinet09}. We can characterise the deformation as a spherical component---which acts as a thermally induced lens, and higher-order effects. At high optical powers, the thermal lens becomes significant and dynamically degrades the mode-matching. The degraded mode-matching has two effects, firstly reducing the optical power in the interferometer and thus reducing the signal-to-noise ratio. Secondly, squeezing losses are increased, limiting the available quantum-enhancement. Additionally, the higher-order effects change the eigenmodes of the resonators. One class of particularly important higher-order effects are point absorbers~\cite{Brooks21}. These absorbers cause localised mirror deformations, significantly changing the mode structure in the resonators. 

In addition to limiting the total achievable squeezing, mode mismatches can coherently scatter light between resonant modes~\cite{Kwee2014}. These modes have a phase relationship that depends on, amongst other things, the accrued Gouy phase. The accrued Gouy phase depends on the thermal lens, which in turn depends on the circulating power. When two points of mismatch exist, resonant and frequency-dependant degradation of the squeezed states can occur~\cite{Kwee2014, toyra17, mcculler20} that vary on thermal timescales. As such, dynamic correction of the mode matching is required. 

In conclusion, quantum enhancement is crucial for achieving the goal 
sensitivities of all detectors, and with ongoing improvements in 
manufacturing and material quality, mode mismatch is becoming the main 
limitation on the level of squeezing.
\section{Beam Profiling and Careful Design}
\label{sec:beam_profiling}
The most direct measure of the beam parameters can be obtained by laser beam profiling. There are two steps: first the beam diameter must be inferred in several positions along the beam axis and second, a least squares fit against \Equation~\ref{eq:beam_radius} must be done to determine the waist size, $w_0$ at a certain position, $z_0$ knowing the wavelength of the radiation denoted as $\lambda$.  
\begin{figure*}
    \centering
    \begin{subfigure}[t]{\linewidth}
         \centering
         \includegraphics[width=.8\linewidth]{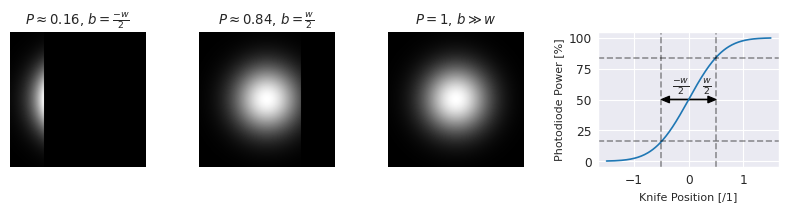}
         \caption{Knife edge technique. The first three sub-plots show a 2D map of beam intensity on the photo-diode, with the knife in different positions. The fourth sub-plot shows \Equation~\ref{eq:knife}, the $x$ axis is in units of beam radius $w$ and the $y$ axis is in percent of maximum power.}
         \label{fig:knife_edge}
     \end{subfigure}\\
    \begin{subfigure}[t]{\linewidth}
         \centering
         \includegraphics[width=.8\linewidth]{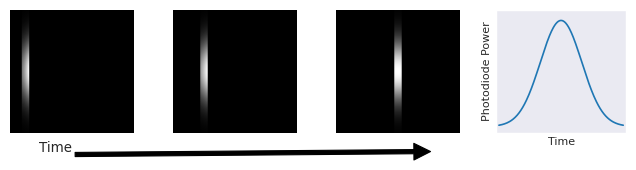}
         \caption{Scanning slit technique. The first three sub-plots show a 2D map of beam intensity on the photo-diode, with the slit in different positions. The fourth sub-plot shows the \Equation~\ref{eq:slit}.}
         \label{fig:slit}
     \end{subfigure}
    \caption{Illustration of \textit{Scanning Slit} and \textit{Knife Edge} techniques to measure a beam width.}
\end{figure*}
The \Gaussian beam intensity profile is described by,
\begin{align}
    I(x,y,z) = \frac{2P}{\pi w^2(z)}\exp \left(\frac{-2(x^2+y^2)}{w^2(z)}\right)
    \tagrepeat{eq:gaussian_beam}
\end{align}
A 1D cross-section is shown in Fig.~\ref{fig:intensity_profile}. We may consider what happens if we block part of the beam with a sharp edge parallel to the $y$ axis,
\begin{align}
    P_k(b) = \int_{y=-\infty}^{y=\infty}\int_{x=-\infty}^b I(x,y,z) = \frac{P}{2}\left[1+\erf{\left(\frac{\sqrt{2}b}{w(z)}\right)}\right].
    \label{eq:knife}
\end{align}%
$P_k(w/2)\approx 0.84$ and $P_k(-w/2)\approx 0.16$. Thus if we slowly block the beam and find the distance between the points where the power is 84\,\% and 16\,\% of its maximum value, we have directly measured the beam radius, $w$. It is important to realise that we are measuring the diameter between $b = \pm w/2$ and the factors of two cancel to yield a radius. This process is called the knife edge technique and is the most traditional method of inferring a beam radius. The scheme is illustrated in \Figure~\ref{fig:knife_edge}.

The slow nature of the measurement means that it is susceptible to beam wander, electrical power drifts and thermal effects. One can improve the measurement precision by using a chopper wheel as a rotating knife edge. If the centre of the beam is distance, $R$, from the choppers centre of rotation then the angle between subtended by the beam is $\theta = w/R$. If the chopper is rotating with linear frequency, $f$, and takes a certain time $\tau$ to travel from the 16\,\% to 84\,\% power levels, then it is trivial to show that
\begin{align}
    w = 2\pi f R \tau.
\end{align}
See~\cite{ThorlabsChopper} for a technical note describing best practices. 

Commercial photodiodes typically have a dynamic range between $10^3$--$10^6$. With careful engineering higher values are possible. This means the above techniques are very suitable for high power beams. However, it is a slow and laborious method of profiling a beam. Furthermore, the measurement assumes a non-astigmatic beam and is insensitive to defects described by higher order modes. Despite working with high power beams regularly, neither the chopper, nor the knife edge technique are frequently used in the gravitational wave-community. Instead, it is very common to use the scanning slit method. One can integrate \Equation~\ref{eq:gaussian_beam} with limits $y=\pm \infty$ and find that the power on the photodiode is,
\begin{align}
    P(x) \approx \sqrt{\frac{2P^2}{\pi}}\exp\left(\frac{-2x^2}{w^2}\right)\frac{\Delta x}{w},
    \label{eq:slit}
\end{align}
where $\Delta x$ denotes the slit width. The approximation is true as long as $\Delta x \ll w$. Devices are available commercially that automatically position the slit, measure the power and fit a Gaussian. Normally, devices are designed with slits in $x$ and $y$ that operate sequentially. In this fashion, the astigmatic beam parameters may be quickly obtained in one measurement, even for very high power beams. This technique is often used for moderate power beams outside the vacuum envelope. However, it is insensitive to effects described by higher order modes. 

Additionally, low power beams are often used for preliminary mode matching. In this instance, it is possible to use a camera with calibrated pixel size to take photos of the beam, \Equation~\ref{eq:gaussian_beam} can then be fit to the intensity map with $w$ and the centroid position as the free parameter. Measurement of astigmatic beams is possible by fitting $w_x$ and $w_y$ independently.  

Regardless of the chosen method, an accurate beam profile requires that the beam width is measured at several points over a distance comparable to the Rayleigh range. From there, it is possible to fit for the two beam parameters $w_0$ and $z_0$. The measurement of beam profiles in this way is common practise in optics labs and the quantum enhanced interferometry community is no different.

\section{Alignment Sensing}
\label{sec:alignment_sensing}
Resonant wavefront sensing is the dominant method of alignment sensing but it is not the only way. For example, lateral effect position sensing was discovered independently by Shottky and Wallmark (as cited in \cite{Henry01}) and has been developed extensively \cite{Yu2010,Wang2021}. Lateral effect sensors have much wider linear regions, which enables new artificial intelligence (AI) alignment and control solutions \cite{Heimann23}.

Quadrant photodiodes (QPD) consist of 4 photo-diodes shaped and arranged symmetrically in a circular or square pattern, as shown in \Figure~\ref{fig:segmented_diodes}. If the alignment of the beam shifts to the right then more light will fall on segments \signal{B} \& \signal{C} than \signal{A} \& \signal{D} and beam pointing signals can be derived. A pitch error signal is given by $V_{\mathrm{pitch}} = A+B-C-D$ and a yaw error signal is given by $V_{\mathrm{yaw}} = B+C-D-A$. These signals can be constructed with a summing amplifier. In the limit that the QPD aperture is much larger than the beam, the beam is Gaussian and offset in pitch/yaw by $b$; then,
\begin{align}
    V_{P/Y} = G(P_k(b) - P_k(-b))
\end{align}
where $P_k$ is given by \Equation~\ref{eq:knife} and $G$ is the product of the amplifier transimpedance gain and responsivity, which is assumed to be the same for all diodes. It is common to normalise $V_{P/Y}$ by the total power $A+B+C+D$ to get a power independent measure of beam alignment. Control of the beam tilt and translation can be achieved by employing this technique twice, as demonstrated by Grafstr\"om \cite{Grafstroem1988}.

In 1984, Anderon showed that a \HG{00} beam, laterally offset at the waist in the $x$ direction by $\Delta x$, may be described as,
\begin{align}
    u_{00}(x&+\Delta x,y) = u_n(x+\Delta x)u_m(y),\\
    &= \left(\frac{2}{\pi}\right)^\frac{1}{4}\sqrt{\frac{1}{w_0}}\exp\left(- \frac{(x+\Delta x)^2}{w^2_0}\right) u_0(y,z),\\
    &=\left(\frac{2}{\pi}\right)^\frac{1}{4}\sqrt{\frac{1}{w_0}}\exp\left(- \frac{x^2}{w^2_0}\right)\exp\left(\frac{2x\Delta x}{w^2_0}\right)\nonumber\\&\quad\exp\left(- \frac{\Delta x^2}{w^2_0}\right) u_0(y).
\end{align}
When $\Delta x \ll w_0$ we may approximate $\exp (\Delta x^2/w^2) \approx 1$ and then Taylor expand $\exp\left(\frac{2x\Delta x}{w^2(z)}\right)$ throwing away terms of $\mathcal{O}(\Delta x^2)$. The result is,
\begin{align}
    u_{00}(x+\Delta x,y) &\approx \left(\frac{2}{\pi}\right)^\frac{1}{4}\sqrt{\frac{1}{w_0}}\exp\left(- \frac{x^2}{w_0^2}\right)\left(1 + \frac{2x\Delta x}{w^2_0}\right) u_0(y),\nonumber\\
    &\approx u_0(x) + \frac{\Delta x}{w_0}u_1(x).
\end{align}
This means an incoming Gaussian laser beam, mismatched to the cavity axis of a laser resonator will scatter light into the HG01 mode. Anderson then developed a scheme to modulate the incoming laser at the cavity mode separation frequency \cite{anderson84}. Thus, the HG01 mode would become resonant in the cavity and hence transmitted by the cavity. Anderson also showed that a Gaussian beam tilted by angle $\theta$ is described by
\begin{align}
    u(x)_\text{tilt} &= u_0(x)\exp\left(\frac{i2\pi x \sin \theta}{\lambda}\right),\\
    &\approx \frac{i\theta}{\theta_\text{div}}u_1(x),
\end{align}
for $\theta_\text{div} = \lambda / (\pi w_0)$. While HG modes are orthogonal when integrated with limits $x,y=\pm \infty$, they are not orthogonal when integrated over areas small with respect to $w(z)$. Thus the transmitted HG01 and HG00 mode will beat on a QPD placed on transmission of the cavity. Conceptually, one can think of this as the cavity converting a static misalignment into a radio frequency beam alignment modulation, which is detected on a QPD at the cavity mode separation frequency. The quadrature of the demodulation changes between detecting translation and tilt misalignments. The scheme was experimentally demonstrated in 1990 \cite{Sampas90}. 
\begin{figure}
    \centering
    \includegraphics[width=\linewidth]{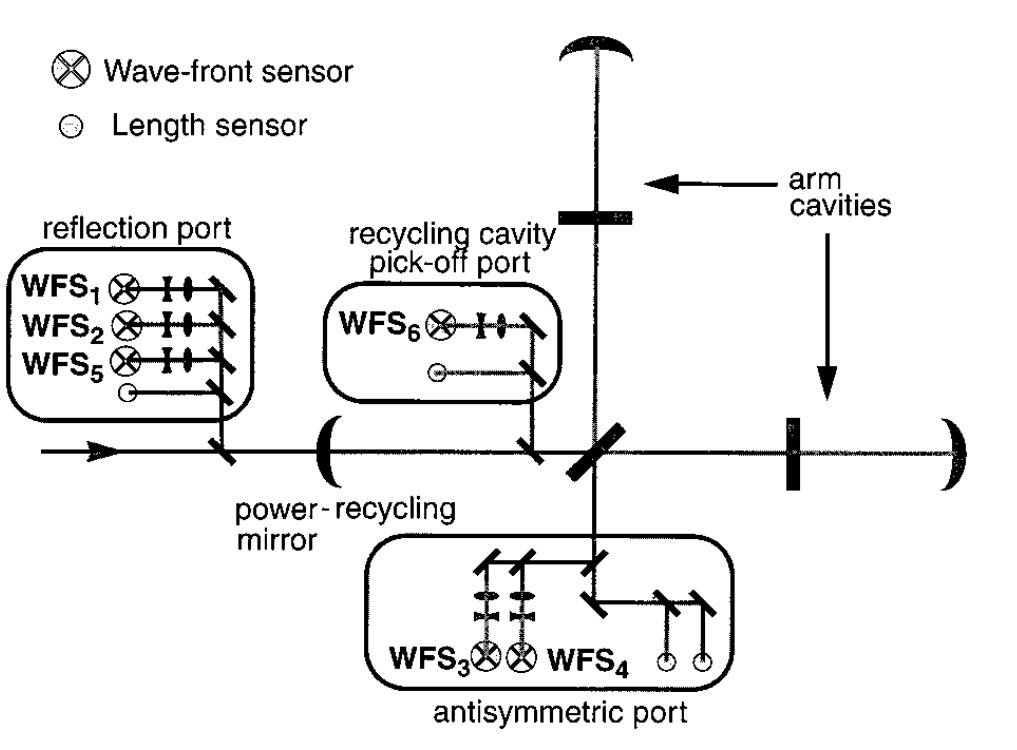}
    \caption{Coupled Cavity alignment sensing scheme. The Ward technique is used at three distinct ports to derive alignment sensing signals for the 10 angular degrees-of-freedom. Reprinted with permission from \cite{Mavalvala98} \textcopyright ~The Optical Society}
    \label{fig:Mavalvala98_fig2}
\end{figure}

In 1994, Morrison \& Ward modified the scheme to work when the laser is modulated at any frequency \cite{Morrison1994,Morrison1994b}. Instead of transmission, non-resonant frequency sidebands are reflected from the cavity along with the non-resonant first-order HG modes. These two modes will beat on a QPD at the modulation frequency. If Pound-Drever-Hall laser frequency stabilisation is in use \cite{drever83,Black01}, those sidebands may be used, reducing the number of frequency modulations. To control both tilt and translation, two QPDs must be used, separated by 90 degrees of accumulated Gouy phase. This scheme is generally referred to as the Ward technique.

The Virgo group summarised and explored a combination of both techniques in \cite{Babusci97}, confirming the presence of suitable error signals in a power-recycled Fabry Perot interferometer \cite{Babusci97}. Between 2004 and 2008, Virgo group used the Anderson scheme with a single sideband providing the length and alignment signals. However, this introduced a length-to-angle coupling, known as the Anderson Offset~\cite{virgo_alog_anderson,virgo_tds_anderson,Matteo_PC}. As the interferometer moved to higher optical powers, heating of the test masses at high power led to a shift of the cavity mode separation frequency, thus shifting the required modulation frequency~\cite{virgo_alog_anderson_high_power}.

Separately to Bayer-Helms, in 1997 Hefetz introduced the BRA-KET notation as a mathematical formalism for tracing beams around complex optical systems \cite{Hefetz1997}. They then use this model to derive the impact of alignment error on the sensitivity of Initial LIGO and then propose a suitable control scheme \cite{Fritschel98}. Later, they perform an experimental test of their proposed alignment scheme \cite{Mavalvala98}. In contrast to \cite{Babusci97} the scheme works entirely in reflection, as shown in \Figure~\ref{fig:Mavalvala98_fig2}. The authors use the term \textit{wavefront sensor} to describe a QPD and associated demodulation circuitry. This scheme was then used in Initial LIGO. In 2005, Slagmolen further explored the technique and published limits an experimentally resized suppression $\sim 50\,dB$ and $\sim 20\,dB$ reductions on low-frequency suspended-optic alignment fluctuations for offset and tilt respectively \cite{Slagmolen2005}.

\end{document}